\documentclass[epj,nopacs]{svjour}
\usepackage{epsfig,array}
\def\nn{\nonumber \\}

\def\fun#1#2{\lower3.6pt\vbox{\baselineskip0pt\lineskip.9pt
\ialign{$\mathsurround=0pt#1\hfil##\hfil$\crcr#2\crcr\sim\crcr}}}

\begin{document}
\newcommand{\be}{\begin{eqnarray}}
\newcommand{\ee}{\end{eqnarray}}
\newcommand{\inli}{\int\limits}


\date{\today}

\title{\bf Partial wave decomposition of
pion and photoproduction amplitudes}

\author
{
A.V.~Anisovich \inst{1,2}
\and E.~Klempt  \inst{1}
\and A.V.~Sarantsev \inst{1,2}
\and U.~Thoma \inst{3}
}

\institute
{
HISKP, Universit\"at Bonn, D-53115 Bonn,
\and Petersburg Nuclear Physics Institute, Gatchina, Russia
\and  Physikalisches Institut, Universit\"at Giessen, Germany
}


\abstract{Partial wave amplitudes for production and
decay of baryon resonances are constructed in the framework of the
operator expansion method. The approach is fully relativistically
invariant and allow us to perform combined analyses
of different reactions imposing directly analyticity and
unitarity constraints. All formulas are given explicitly in the form
used by the Crystal Barrel collaboration in the (partly forthcoming)
analyses of the
electro-, photo- and pion induced meson production data.
}




\titlerunning{\bf Partial wave decomposition of
pion and photoproduction amplitudes}

\mail{klempt@hiskp.uni-bonn.de}

\maketitle

\section{Introduction}

The perturbative approach to the theory of strong interaction
(perturbative QCD) cannot be applied directly to the region of
low and intermediate energies. In spite of many efforts to create
a non-perturbative formulation for QCD from first principles, a final
breakthrough has not yet been achieved even if recent results of
lattice QCD indicate that this situation might change in the future. A
necessary step towards
a better understanding of strong interactions is undoubtedly a
precise knowledge of the experimental situation and a correct
classification of strongly interacting particles.

In meson spectroscopy, considerable progress had been made
during the last ten years. A variety of experiments lead to the
discovery of a large number of new meson states. In particular
scalar states, very poorly known 15 years ago, are now one
of the most studied systems. As a result, it is now possible to
investigate systematically the question if additional states
expected from QCD like glueballs or hybrids hide in the observed
meson spectrum. Although there is still no agreement on the
classification of scalar states, the number of reliable
classifications is reduced to quite a small number (see
\cite{close_cl,anisov_cl,minkochs,klempt_cl,Tornqvist} and references
therein). We expect that the new GSI facility will help to resolve
the remaining ambiguities completely.

A very important observation is that those meson resonances
which can be interpreted as dominant $q\bar q$ states are lying on
linear trajectories, not only against the total spin but also against
their radial quantum number \cite{anis_tr}. Excitingly, this seems to
be true also for baryons \cite{klempt_tr}. Almost all known baryons
lye on linear trajectories with the same slope as that for mesons.

Most information about baryons comes from pion- and photon-induced
production of single mesons. However the experience from meson
spectroscopy shows that excited states decay dominantly into
multi-body channels and are not observed reliably in the elastic
cross section. Thus reactions with three or more final states provide
rich information about the properties of hadronic resonances. One of
recent examples is the possible observation of a pentaquark \cite{klempt_cl}
which up to now was seen only in reactions with three or more
final-state particles.

The task to extract pole positions and residues from multi-body final
states is however not a simple one. Main problems can be traced to the
large interference effects between different isobars and to
contributions from singularities related to multi-body interactions. In
\cite{Anisovich:sing} an approach based on the dispersion N/D method
was put forward and successfully applied to the analysis of meson
resonances. In this method  singularities in the
reaction can be classified, resonances which are closest to
the physical region can be taken into account accurately.
Other contributions can be parameterized in an efficient way.

One of the key points in this approach is the operator decomposition
method which provides a tool for a universal construction of
partial wave amplitudes for reactions with two-- and many--body final
states. The operator decomposition method has a long history. It was
used for the analysis of the reactions with three particle final states
already in \cite{anis_daxno}. The full description of the method for
the nonrelativistic case was given in \cite{zemach}. The full
relativistic approach for $NN\to NN (N\Delta)$ and $\gamma d \to pn$
was developed in \cite{oper_1,oper_2,oper_3}. The construction of
partial wave amplitudes for production of meson resonances in
different reactions can be found in \cite{chung,xoper,oper_gg}.

In the present article we develop the operator expansion
method to describe baryon resonances in meson- and photon-induced
reactions.  The photon can be real or virtual, we assume it to
be virtual unless the opposite is explicitly stated. The method is also
very convenient to calculate contributions from triangle and box
diagrams and to project t and u-channel exchange amplitudes
into partial waves. The latter feature is very important for
amplitudes near their unitarity limits where the unitarity property
must be taken into account explicitly.

The formulas given here reproduce exactly the amplitudes
used by the Crystal-Barrel-ELSA collaboration in their
(partly forthcoming) analyses of single- and two-body
photoproduction reactions.

It must be emphasized that a wealth of data on baryon resonances
has been taken, is being analyzed or is going to be produced in the
near future. At MAMI in Mainz~\cite{MAMI} precision data were taken in
the low--energy range which will be extended to 1.47\,GeV photon
energies in the close future. The GRAAL~\cite{GRAAL} experiment has
produced  invaluable data in particular using linearly polarized
photons; the SAPHIR~\cite{SAPHIR} experiment at Bonn has published
a series of papers covering many basic photoproduction cross
sections. The experiment is now replaced by the Crystal Barrel
detector~\cite{CBAR} which had produced before many results at the
Low-Energy-Antiproton-Ring (LEAR) at CERN. And, last not least, Jlab
at Newport News/Virginia ~\cite{JLAB} has accumulated high statistic
data sets on photo- and electro-production of a variety of final states. First
high--quality data have been published.

\subsection{Orbital angular momentum operators }

Let us consider a decay of a composite particle with
spin $J$ and momentum $P$ ($P^2=s$) into two spinless particles with
momenta $k_1$ and $k_2$. The only measured quantities in such a reaction
are the particle momenta. The angular dependent part of
the wave function of the composite state is described by operators
constructed out of these momenta and the metric tensor. Such operators
(we will denote them as $X^{(L)}_{\mu_1\ldots\mu_L}$, where $L$ is the
orbital momentum)
are called orbital angular momentum operators and
correspond to irreducible representations of the Lorentz group.
They satisfy the following properties \cite{xoper}:
\begin{itemize}
\item Symmetry with respect to permutation of any two indices:
\be
X^{(L)}_{\mu_1\ldots\mu_i\ldots\mu_j\ldots\mu_L}\; =\;
X^{(L)}_{\mu_1\ldots\mu_j\ldots\mu_i\ldots\mu_L}.
\label{oth_b1}
\ee
\item Orthogonality to the total momentum of the system, $P=k_1+k_2$.
\be
P_{\mu_i}X^{(L)}_{\mu_1\ldots\mu_i\ldots\mu_L}\ =\ 0.
\label{oth_b2}
\ee
\item The traceless property for summation over two any indices:
\be
g_{\mu_i\mu_j}X^{(L)} _{\mu_1\ldots\mu_i\ldots\mu_j\ldots\mu_L}\
\ =\ 0.
\label{oth_b3}
\ee
\end{itemize}

Let us consider a one-loop diagram describing the decay of a composite
system into two spinless particles which propagate and then form again
a composite system. The decay and formation processes are described by
orbital angular momentum operators. Due to conservation of quantum
numbers this amplitude must vanish for initial and final states with
different spin. The S-wave operator is a scalar and can be taken as
unit operator. The P-wave operator is a vector. In the dispersion relation
approach it is sufficient that the imaginary part of the loop
diagram with S and P-wave operators as vertices is equal to 0.
In the case of spinless particles this requirement entails
\be
\int\frac{d\Omega}{4\pi} X^{(1)}_\mu =0.
\ee
where the integral is taken over the solid angle of the
relative momentum. In general the result of such an integration
is proportional to the total momentum of the system $P_\mu$ (the only
external vector):
\be
\int\frac{d\Omega}{4\pi} X^{(1)}_\mu =\lambda P_\mu\;.
\ee
Convoluting this expression with $P_\mu$ and demanding $\lambda=0$
we obtain the orthogonality condition (\ref{oth_b2}).
The orthogonality between D-wave and S-wave is provided by the
traceless condition (\ref{oth_b3}); conditions
(\ref{oth_b2},\ref{oth_b3}) provide the orthogonality for all
operators with different orbital angular momenta.

The orthogonality condition (\ref{oth_b2})
is automatically fulfilled if the operators are
constructed from the relative momenta
$k^\perp_\mu$ and the tensor $g^\perp_{\mu\nu}$. They both are
orthogonal to the total momentum of the system:
\be
k^\perp_\mu=\frac12
g^\perp_{\mu\nu}(k_1-k_2)_\nu  \;; \qquad\qquad
g^\perp_{\mu\nu}=g_{\mu\nu}-\frac{P_\mu P_\nu}{s}\;.
\ee
In the center-of-mass system (c.m.s. from now onwards),
where $P=(P_0,\vec P)=(\sqrt s,0)$, the vector
$k^\perp$ is space-like: $k^\perp=(0,\vec k)$.

The operator for $L=0$ is a scalar (for example a unit
operator),
and the operator for $L=1$ is a vector which can only be
constructed from $k^\perp_\mu$.
The orbital angular momentum operators for $L =0 $ to 3
are:
\begin{eqnarray}
X^{(0)}&=&1\ , \qquad X^{(1)}_\mu=k^\perp_\mu\ , \qquad\nonumber \\
X^{(2)}_{\mu_1 \mu_2}&=&\frac32\left(k^\perp_{\mu_1}
k^\perp_{\mu_2}-\frac13\, k^2_\perp g^\perp_{\mu_1\mu_2}\right), \nonumber  \\
X^{(3)}_{\mu_1\mu_2\mu_3}&=& \\
\frac52\Big[k^\perp_{\mu_1} k^\perp_{\mu_2 }
k^\perp_{\mu_3} &-&
\frac{k^2_\perp}5\left(g^\perp_{\mu_1\mu_2}k^\perp
_{\mu_3}+g^\perp_{\mu_1\mu_3}k^\perp_{\mu_2}+
g^\perp_{\mu_2\mu_3}k^\perp_{\mu_1}
\right)\Big] . \nonumber
\ee
The operators $X^{(L)}_{\mu_1\ldots\mu_L}$ for $L\ge 1$ can be written in
form of a recurrent expression:
\be
X^{(L)}_{\mu_1\ldots\mu_L}&=&k^\perp_\alpha
Z^{\alpha}_{\mu_1\ldots\mu_L} \; ,
\nonumber\\
Z^{\alpha}_{\mu_1\ldots\mu_L}&=&
\frac{2L-1}{L^2}\Big (
\sum^L_{i=1}X^{{(L-1)}}_{\mu_1\ldots\mu_{i-1}\mu_{i+1}\ldots\mu_L}
g^\perp_{\mu_i\alpha}-
\nonumber \\
 \frac{2}{2L-1}  \sum^L_{i,j=1 \atop i<j}
&g^\perp_{\mu_i\mu_j}&
X^{{(L-1)}}_{\mu_1\ldots\mu_{i-1}\mu_{i+1}\ldots\mu_{j-1}\mu_{j+1}
\ldots\mu_L\alpha} \Big ).
\label{z}
\ee
Convolution equality reads:
\be
X^{(L)}_{\mu_1\ldots\mu_L}k^\perp_{\mu_L}=k^2_\perp
X^{(L-1)}_{\mu_1\ldots\mu_{L-1}}.
\label{ceq}
\ee
Based on eq.(\ref{ceq}) and
taking into account the traceless property of $X^{(L)}_{\mu_1\ldots\mu_L}$,
one can write down the orthogonality-normalization condition for
orbital angular  operators:
\be
\int\frac{d\Omega}{4\pi}
X^{(n)}_{\mu_1\ldots\mu_n}(k^\perp)X^{(m)}_{\mu_1\ldots\mu_m}(k^\perp)
=\delta_{nm}\alpha(L)k^{2n}_\perp \; , \nonumber
\\
\alpha(L)=\prod^L_{l=1}\frac{2l-1}{l}=\frac{(2L-1)!!}{L!} \; .
\label{alpha}
\ee
Iterating eq. (\ref{z}) one obtains the
following expression for the operator $
X^{(L)}_{\mu_1\ldots\mu_L}$:
\be
X^{(L)}_{\mu_1\ldots\mu_L}(k^\perp)
=\alpha(L)
\bigg [
k^\perp_{\mu_1}k^\perp_{\mu_2}k^\perp_{\mu_3}k^\perp_{\mu_4}
\ldots k^\perp_{\mu_L}
&-&  \nonumber \\
\frac{k^2_\perp}{2L-1}\bigg(
g^\perp_{\mu_1\mu_2}k^\perp_{\mu_3}k^\perp_{\mu_4}\ldots
k^\perp_{\mu_L} &+& \nonumber \\
g^\perp_{\mu_1\mu_3}k^\perp_{\mu_2}k^\perp_{\mu_4}\ldots
k^\perp_{\mu_L} &+& \ldots \bigg)
+ \\
\frac{k^4_\perp}{(2L\!-\!1)
(2L\!-\!3)}\bigg(
g^\perp_{\mu_1\mu_2}g^\perp_{\mu_3\mu_4}k^\perp_{\mu_5}
k^\perp_{\mu_6}\ldots k_{\mu_L} &+&
\nonumber \\
g^\perp_{\mu_1\mu_2}g^\perp_{\mu_3\mu_5}k^\perp_{\mu_4}
k^\perp_{\mu_6}\ldots k_{\mu_L}&+&
\ldots\bigg)+\ldots\bigg ].  \nonumber
\label{x-direct}
\ee

When a composite system decays into two spinless particles
the total spin
is defined by the angular momentum only ($J=L$) and
the angular part of the scattering amplitude
(for example a $\pi\pi\to\pi\pi$ transition) is described
as a convolution of the operators
$X^{(L)}(k)$ and $X^{(L)}(q)$ where $k$ and $q$ are relative
momenta before and after the interaction.
\be
X^{(L)}_{\mu_1\ldots\mu_L}(k^\perp)X^{(L)}_{\mu_1\ldots\mu_L}
(q^\perp) = &&  \\ \alpha(L)
&& \left(\sqrt{k^2_\perp}\sqrt{ q^2_\perp}\right)^{L} P_L(z)\; .
\nonumber
\label{kq_ampl}
\ee
Here $P_L(z)$ are Legendre polynomials (see Appendix A) and
$z=(k^\perp q^\perp)/(\sqrt{k_\perp^2}\sqrt{q_\perp^2})$ which are,
in c.m.s., functions of the cosine of the angle
between initial and final particles.\\
A comment: one should be careful with expression
$\sqrt{k_\perp^2}$. In c.m.s.,
\be
&&\sqrt{k_\perp^2}=\sqrt{-\vec k^2}=i|\vec k|\;,
\nonumber \\
&&(\sqrt{k_\perp^2}\sqrt{q_\perp^2})^{L}=
(-1)^L(|\vec k||\vec q|)^L \;.
\ee

\subsection{The boson projection operator}

Let us consider the imaginary part of the one-loop diagram when
particles interact with relative momentum $p$, then propagate with
momentum $k$, and interact for a second time getting the
relative momentum
$q$. The process can be described by orbital
angular momentum operators in the form
\be
X^{(L)}_{\mu_1\ldots\mu_L}(p^\perp)
\int\frac{d\Omega }{4\pi}\
X^{(L)}_{\mu_1\ldots\mu_L}(k^\perp)
X^{(L)}_{\nu_1\ldots\nu_L}(k^\perp)
X^{(L)}_{\nu_1\ldots\nu_L}(q^\perp)   \nonumber
\ee
The projection operator $O^{\mu_1\ldots\mu_L}_{\nu_1\ldots \nu_L}$
for the partial wave with angular momentum $L$ is defined as:
\be
\int\frac{d\Omega }{4\pi}\
X^{(L)}_{\mu_1\ldots\mu_L}(k^\perp)
X^{(L)}_{\nu_1\ldots\nu_L}(k^\perp) & = & \nonumber \\
 \frac{\alpha(L)}{2L+1} & k^{2L}_\perp &
O^{\mu_1\ldots\mu_L}_{\nu_1\ldots \nu_L}
\label{18}
\ee
and satisfies the following relations:
\be
X^{(L)}_{\mu_1\ldots\mu_L}(k^\perp)
O^{\mu_1\ldots\mu_L}_{\nu_1\ldots \nu_L}\
=\ X^{(L)}_{\nu_1\ldots \nu_L}(k^\perp)\ , \nonumber \\
O^{\mu_1\ldots\mu_L}_{\alpha_1\ldots\alpha_L} \
O^{\alpha_1\ldots\alpha_L}_{\nu_1\ldots \nu_L}\
=\ O^{\mu_1\ldots\mu_L}_{\nu_1\ldots \nu_L}\ .
\label{proj_op}
\ee

Due to properties (\ref{proj_op}), the product of
any number of
loop diagrams will be described by the same projection operator.
This operator has the same symmetry, orthogonality and traceless
properties as $X$-operators (for the same set
of up and down indices) but the $O$-operator
 does not depend on the relative momentum
of the constituents and does not describe decay
processes. It represents the propagation of the composite system and
defines the structure of the boson
propagator (its numerator). More details on the
properties of $X$ and $O$-operators can be found in \cite{xoper}.

Taking into account the definition of the projection operators
(\ref{proj_op})
and the properties of the $X$-operators (\ref{x-direct}) we obtain:
\be
k_{\mu_1}\ldots k_{\mu_L}
O^{\mu_1\ldots\mu_L}_{\nu_1\ldots \nu_L}\ =
\frac{1}{\alpha(L)}
X^{(L)}_{\nu_1\ldots\nu_L}(k^\perp).
\label{19}
\ee
This equation presents the basic property of
the projection operator:
it projects any operator with $L$ indices onto the partial
wave operator with angular momentum $L$. \\
The projection operator can also be calculated using the recurrent
expression:
\be
&&O^{\mu_1\ldots\mu_L}_{\nu_1\ldots \nu_L}=\frac{1}{L^2}
\bigg ( \sum\limits_{i,j=1}^{L}g^\perp_{\mu_i\nu_j}
O^{\mu_1\ldots\mu_{i-1}\mu_{i+1}\ldots\mu_L}_{\nu_1\ldots
\nu_{j-1}\nu_{j+1}\ldots\nu_L}-
\nonumber \\
 &&  \frac{4}{(2L-1)(2L-3)} \\    &&
\sum\limits_{i<j\atop k<m}^{L}
g^\perp_{\mu_i\mu_j}g^\perp_{\nu_k\nu_m}
O^{\mu_1\ldots\mu_{i-1}\mu_{i+1}\ldots\mu_{j-1}\mu_{j+1}\ldots\mu_L}_
{\nu_1\ldots\nu_{k-1}\nu_{k+1}\ldots\nu_{m-1}\nu_{m+1}\ldots\nu_L}
\bigg )   \nonumber
\ee
The low order projection operators are:
\be
O=1 \;,\qquad\qquad O^\mu_\nu=g^\perp_{\mu\nu} \;,
\nonumber \\
O^{\mu\nu}_{\alpha\beta}=\frac 12 \Big (
g^\perp_{\mu\alpha}g^\perp_{\nu\beta}+
g^\perp_{\mu\beta}g^\perp_{\nu\alpha}-\frac 23
g^\perp_{\mu\nu}g^\perp_{\alpha\beta} \Big ) \;.
\ee

\subsection{The vector projection operator in the gauge invariant
limit}

The sum over the  possible polarisations of a vector
particle $\varepsilon_\mu$ with non zero mass
corresponds to the vector projection operator:
\be
\sum\limits_\alpha \varepsilon^{\alpha}_\mu\varepsilon^{*\alpha}_\nu=
O^\mu_\nu=g^\perp_{\mu\nu}\;.
\label{proj_vec}
\ee
which means that there are three independent polarisation vectors
orthogonal to the momentum of the particle and normalized as
$\varepsilon^\alpha_\mu\varepsilon^{*\alpha}_\mu=-1$.

However photon polarisation vectors have only two independent components,
its momentum squared is equal to 0 and
therefore, the projection operator can not have the form
(\ref{proj_vec}). The invariant expression for the photon projection
operator can be only constructed for the interaction of the photon with
another particle. In this case it has the form:
\be
g_{\mu\nu}^{\perp\perp}=
-\sum\limits_{\alpha}
\varepsilon^{\alpha}_\mu
\varepsilon^{\alpha}_\nu=
g_{\mu\nu}-\frac{P_\mu P_\nu}{P^2}-\frac{
k^\perp_\mu k^\perp_\nu}{k_\perp^2}\;.
\label{g_2p}
\ee
where $k_1$ is momentum of the baryon, $k_2$ -is momentum
of the photon, $P=k_1+k_2$
and
\be
k^\perp_\mu=\frac 12 (k_1-k_2)_\nu g^\perp_{\mu\nu}=
\frac 12 (k_1-k_2)_\nu \left ( g^\perp_{\mu\nu}-\frac{P_\mu P_\nu}{P^2}
\right )
\ee
In the c.m.s. with the momentum of $\gamma$ being parallel to the z-axis,
the $g^{\perp\perp}_{\mu\nu}$ tensor has a very simple form:
\be
g^{\perp\perp}_{\mu\nu}\qquad = \qquad
\left (
\begin{array}{cccc}
0 & 0 & 0 &0 \\
0 &-1 & 0 &0 \\
0 & 0 &-1 &0 \\
0 & 0 & 0 &0
\end{array}
\right )
\ee
where the vector  components are defined as \\ $p=(E,p_x,p_y,p_z)$.

The tensor (\ref{g_2p}) is orthogonal to the momentum of the both
particles:
\be
g_{\mu\nu}^{\perp\perp} k_{2\mu}=
g_{\mu\nu}^{\perp\perp} k_{1\mu}=0\;.
\ee
and it extracts the gauge invariant
part of the amplitude. For the real photon:
\be
A=A_\mu \varepsilon^\alpha_\mu=
A_\nu g^{\perp\perp}_{\nu\mu}\varepsilon^\alpha_\mu \;.
\ee
and the expression $A_\nu g^{\perp\perp}_{\nu\mu}$ is gauge invariant:\\
$A_\nu g^{\perp\perp}_{\nu\mu} k_{2\mu}=0$.

\section{Fermions}

The wave function of a fermion is described as Dirac $\;$
bispinor, as object in Dirac space represented by
$\gamma$ matrices.
In the standard representation the $\gamma$ matrices have
the following form:
\be
\gamma_0=\left
( \begin{array}{cc} 1 & 0 \\ 0 & -1 \end{array}
\right ),\qquad
\vec \gamma=\left (
\begin{array}{cc}
0 & \vec \sigma \\
-\vec \sigma & 0
\end{array}
\right ),\qquad
\gamma_5=\left (
\begin{array}{cc}
0 & 1 \\
1 & 0
\end{array}
\right )\qquad
\ee
where $\vec \sigma$ are $2\times 2$ Pauli matrices.
In this representation the spinors for fermion particles with momentum p
are:
\be
u(p)=\frac{1}{\sqrt{p_0+m}}
\left (
\begin{array}{c}
(p_0+m)\omega \\
(\vec p\vec \sigma)\omega
\end{array}
\right ), \nonumber \\
\bar u(p)=\frac{
\left ( \omega^* (p_0+m),
-\omega^*(\vec p\vec \sigma) \right )
}{\sqrt{p_0+m}}.
\ee
Here $\omega$ represents a 2-dimensional spinor
and $\omega^*$ the conjugated and transposed spinor. The normalization
condition can be written as:
\be \bar u(p) u(p)=2m  \quad
\sum\limits_{polarizations} u(p) \bar u(p)=m+\hat p
\label{bisp_norm}
\ee
We define $\hat p =p^\mu\gamma_\mu$.

\section{The structure of fermion propagator}

The wave function of a particle with spin $J=L+1/2$ and momentum $p$
is described by a tensor bispinor $\Psi_{\mu_1\ldots\mu_L}$: it is a
tensor in Dirac space.
As the tensor it
satisfies the same properties as a boson wave function:
\be &
p_{\mu_i}\Psi_{\mu_1\ldots\mu_L}&=0\;,
\nonumber \\ &
\Psi_{\mu_1\ldots\mu_i\ldots\mu_j\ldots\mu_L}&=
\Psi_{\mu_1\ldots\mu_j\ldots\mu_i\ldots\mu_L}\;,
\nonumber \\
&  g_{\mu_i\mu_j}\Psi_{\mu_1\ldots\mu_L}&=0 \;.
\label{fprop_1}
\ee
In addition the fermion wave function must satisfy the
following properties:
\be
&(\hat p-m)\Psi_{\mu_1\ldots\mu_L}&=0 \;,
\nonumber \\
&\gamma_{\mu_i}\Psi_{\mu_1\ldots\mu_L}&=0 \;.
\label{fprop_2}
\ee
Conditions (\ref{fprop_1}), (\ref{fprop_2}) define the
structure of the fermion propagator  (projection operator) which can be
written in the following form:
\be
F^{\mu_1\ldots\mu_L}_{\nu_1\ldots
\nu_L}(p)=
(m+\hat p) R^{\mu_1\ldots\mu_L}_{\nu_1\ldots \nu_L}
\label{fp_1st}
\ee
Here $(m+\hat p)$ corresponds to the propagator for a fermion with
$J=1/2$.\\
The operator $R^{\mu_1\ldots\mu_L}_{\nu_1\ldots \nu_L}$ describes
the tensor
structure of the propagator. It is equal to 1 for a $J=1/2$ particle
and is proportional to
$g^\perp_{\mu\nu}-\gamma^\perp_\mu\gamma^\perp_\nu/3$ for a
particle with spin $J=3/2$ ($\gamma^\perp_\mu=g^\perp_{\mu\nu}\gamma_\nu$).

The conditions (\ref{fprop_1}) are identical for fermion and boson
projection operators and therefore the fermion
projection operator can be written as:
\be
R^{\mu_1\ldots\mu_L}_{\nu_1\ldots \nu_L}=
O^{\mu_1\ldots\mu_L}_{\alpha_1\ldots \alpha_L}
T^{\alpha_1\ldots\alpha_L}_{\beta_1\ldots \beta_L}
O^{\beta_1\ldots \beta_L}_{\nu_1\ldots\nu_L}
\ee
The $T^{\alpha_1\ldots\alpha_L}_{\beta_1\ldots \beta_L}$ operator
can be expressed in a rather simple
form since all symmetry and orthogonality conditions
are  imposed by  $O$-operators.
First, the T-operators are constructed only out
of metrical tensors and $\gamma$-matrices.
Second, a construction like
$\gamma_{\alpha_i}\gamma_{\alpha_j}$
\be
\gamma_{\alpha_i}\gamma_{\alpha_j}=
\frac12 g_{\alpha_i\alpha_j}+\sigma_{\alpha_i\alpha_j},  \\
{\rm where} \qquad
\sigma_{\alpha_i\alpha_j}=\frac 12
(\gamma_{\alpha_i}\gamma_{\alpha_j}-\gamma_{\alpha_j}\gamma_{\alpha_i}) \nonumber
\ee
gives zero if multiplied with an
$O$-operator (the first term due to
the traceless conditions and the second one due to symmetry properties).
The only structures which can then be constructed
are $g_{\alpha_i\beta_j}$ and $\sigma_{\alpha_i\beta_j}$. Moreover,
taking into account the symmetry properties of the $O$-operators, the latter
can be used as $\sigma_{\alpha_1\beta_1}$:
\be
T^{\alpha_1\ldots\alpha_L}_{\beta_1\ldots \beta_L}=
\frac{L+1}{2L\!+\!1}
\big( g_{\alpha_1\beta_1}-
\frac{L}{L\!+\!1}\sigma_{\alpha_1\beta_1} \big)
\prod\limits_{i=2}^{L}g_{\alpha_i\beta_i}
\label{t1}
\ee
Here the coefficients are calculated to satisfy the
conditions (\ref{fprop_2})
for the fermion projection operator:
\be
\gamma_{\mu_i}F^{\mu_1\ldots\mu_L}_{\nu_1\ldots \nu_L}
=F^{\mu_1\ldots\mu_L}_{\nu_1\ldots \nu_L}\gamma_{\nu_j}=0\;,
\\
F^{\mu_1\ldots\mu_L}_{\alpha_1\ldots \alpha_L}
F^{\alpha_1\ldots \alpha_L}_{\nu_1\ldots \nu_L}=
F^{\mu_1\ldots\mu_L}_{\nu_1\ldots \nu_L} \;.
\ee
It is not necessary  to construct the
T operator out of the metric tensors and
$\sigma$-matrices
orthogonal to the momentum of the particle. Orthogonality
is imposed
by $O$-operators. However, to use the same ingredients
for all operators, it is easier to introduce this property directly,
rewriting the T-operators as:
\be
T^{\alpha_1\ldots\alpha_L}_{\beta_1\ldots \beta_L}&=&
\frac{L+1}{2L\!+\!1}
\big( g^\perp_{\alpha_1\beta_1}-
\frac{L}{L\!+\!1}\sigma^\perp_{\alpha_1\beta_1} \big)
\prod\limits_{i=2}^{L}g^\perp_{\alpha_i\beta_i}, \\
\sigma^\perp_{\mu\nu}&=&\frac 12
(\gamma^\perp_{\mu}\gamma^\perp_{\nu}-
\gamma^\perp_{\nu}\gamma^\perp_{\mu})  \nonumber
\label{t2}
\ee

\subsection{Fermion propagator for an unstable particle}

The numerator of a stable particle
propagator has a very simple structure
in its c.m.s.:
\be
m +\hat P =2m
\left ( \begin{array}{cc} 1 & 0 \\ 0 & 0 \end{array} \right ).
\ee

Assume a resonance with an invariant mass $\sqrt{s}$ ($P^2=s$).
To maintain the orthogonality condition for the operators one should
replace $m\to \sqrt {s}$ in eq. (\ref{fp_1st}).
Then, for a resonance in its c.m.s.:
\be
\sqrt{s} +\hat P =2\sqrt{s}
\left ( \begin{array}{cc} 1 & 0 \\ 0 & 0 \end{array} \right ).
\ee

Such a structure is divergent at large energies
and it is reasonable to regularize it with the factor
$2M/(2\sqrt s)$ or simply with $1/(2\sqrt s)$ to
provide a correct asymptotical behavior.
Therefore we use the following
expression for the numerator of a resonance propagator:
\be
F^{\mu_1\ldots\mu_L}_{\nu_1\ldots
\nu_L}(P)=
\frac{\sqrt{s}+\hat P}{2\sqrt s} R^{\mu_1\ldots\mu_L}_{\nu_1\ldots \nu_L}\;.
\label{fp_res}
\ee

\section{\boldmath $\pi N$ \unboldmath scattering}

Let us now construct vertices for the decay of a composite baryon
system with momentum $P$ into the $\pi N$ final state with relative
momentum $k=1/2(k_1-k_2)$ (here $k_1$ -is the nucleon momentum).
A particle with spin $J^P=1/2^-$ decays into the $\pi N$ channel
in an S-wave, hence the orbital angular momentum
operator is a scalar, e.g. a unit operator. For the vertex we get:
\be
\bar u(P) u(k_1).
\ee
Here $u(P)$ is a bispinor of the composite particle and
$u(k_1)$ is the bispinor of the nucleon. A resonance
with spin $3/2^+$ decays into $\pi N$ with an orbital angular
momentum $L=1$ and the vertex must be a vector,
constructed out of $k^\perp_\mu$ and $\gamma^\perp_\mu$. However it is
sufficient to take only $k^\perp_\mu$: first, due to the properties
(\ref{fprop_2}) and second, due to the fact that the projection
operator (numerator of the fermion propagator)
will automatically provide the correct structure.
Thus we obtain for the decay of particles with
$J=(L+1/2)$, $P=(-1)^{L+1}$ ($1/2^-$, $3/2^+$, $5/2^-$, $7/2^+$,
$\ldots$) the expression
\be
\bar \Psi_{\mu_1\ldots\mu_L} X^{(L)}_{\mu_1\ldots\mu_L}(k^\perp)
u(k_1).
\ee
Let us call this set of states where
the total angular momentum is given by
the orbital angular momentum plus $1/2$ as 'plus' or '+' states.
'Minus' or '-' states are defined analogously ($J=(L-1/2)$, $P=(-1)^{L+1}$).

It is convenient to introduce vertex functions
$N^\pm_{\mu_1\ldots\mu_L}$
describing the decay of a resonance into a pseudoscalar meson and a
nucleon. Then for '+' states:
\be
\bar \Psi_{\mu_1\ldots\mu_L}
N^+_{\mu_1\ldots\mu_L}(k^\perp) u(k_1) \;, \nonumber \\
N^+_{\mu_1\ldots\mu_L}(k^\perp)=
X^{(L)}_{\mu_1\ldots\mu_L}(k^\perp)\;.
\ee \\
The angular dependent part of the $\pi N\to {\rm resonance } \to \pi N$
transition amplitude can be con\-structed in a very simple way:
the vertex function describing the interaction of the meson and
the nucleon convolutes with the intermediate state
propagator and the decay vertex function:
\be
\bar u_f \tilde N^\pm_{\mu_1\ldots\mu_L}
F^{\mu_1\ldots\mu_L}_{\nu_1\ldots\nu_L}(P)
N^\pm_{\nu_1\ldots\nu_L} u_i
\ee
Here $\tilde N^\pm$ is the left-hand vertex function
(with two particles joining to one resonance)
which is different from the decay vertex function $N^\pm$ by
the ordering of $\gamma$-matrices.
(This is important for $N^-_{\mu_1\ldots\mu_L}$ vertices
which will be given on the next page).
If $q$ and $k$ are the relative momenta
before and after interaction and $k_1$ and $q_1$ are the
corresponding nucleon momenta, the amplitude for $\pi N$ scattering via
'+' states can be written as:
\be
A &= &\\  & \bar u(k_1)& X_{\mu_1\ldots\mu_L}(k^\perp)
F^{\mu_1\ldots\mu_L}_{\nu_1\ldots\nu_L}(P)
X_{\nu_1\ldots\nu_L}(q^\perp)
u(q_1) BW_L^+(s) \nonumber
\ee
where $BW_L^+(s)$ describes the energy dependence of the intermediate
state propagator. It is given, e.g., by a Breit-Wigner amplitude, a
K-matrix or an N/D expression.

Using equations (\ref{proj_op}) and (\ref{t2}) we obtain:
\be
&A=&\bar u(k_1)\frac{\sqrt s+\hat P}{2\sqrt s} u(q_1)
(\sqrt{k_\perp^2}\sqrt{q_\perp^2})^L
\frac{L+1}{2L\!+\!1} \alpha(L)\;  \nonumber \\
&& P_L(z)BW_L^+(s)-
\bar u(k_1)\frac{\sqrt s+\hat P}{2\sqrt s}\frac{L}{2L\!+\!1}
\sigma^\perp_{\mu\nu}  \\ &&
X_{\mu\mu_2\ldots\mu_L}(k^\perp)
X_{\nu\mu_2\ldots\mu_L}(q^\perp) u(q_1) BW_L^+(s) \;.
\nonumber
\ee

The formulas for the convolution of $X$-operators
with one free index in each operator is given in Appendix B
eq.(\ref{x11}).
Only the last, antisymmetric term, gives a nonzero result:
\be
&A=&
(\sqrt{k_\perp^2}\sqrt{q_\perp^2})^L
\bar u(k_1)\frac{\sqrt s+\hat P}{2\sqrt s}
\frac{\alpha(L)}{2L\!+\!1}BW_L^+(s)
\nonumber\\ &&
\Big [ (L+1)P_L(z)-
\frac{\sigma_{\mu\nu}k_\mu q_\nu}{\sqrt{k_\perp^2}\sqrt{q_\perp^2}}
P'_L(z)\Big ] u(q_1) \;.
\label{12m}
\ee

Let us now construct the vertices for the decay of composite particles
with spin-parity $1/2^+,3/2^-,5/2^+\ldots$ into $\pi N$.
The state
with $1/2^+$ is a scalar in tensor space and
decays into $\pi N$ with $L=1$. Therefore this scalar should
be constructed from $k^\perp_\mu$.
It cannot be $\hat k^\perp=k^\perp_\mu\gamma_\mu$ since
such an operator is not orthogonal to the $1/2^-$ state:
\be
\bar u(P) \hat k^\perp u(k_1) &=&
\bar u(P) (\hat k_1-\alpha \hat P) u(k_1)= \nonumber \\ &&
\bar u(P) u(k_1) (m_1-a(s) \sqrt s)\;.
\label{decomp}
\ee
Here we used:
\be
k_{1\mu}=k^\perp_\mu+a(s) P_\mu, \\
{\rm with}\quad a(s)=\frac{Pk_1}{P^2}=\frac{s+m_N^2-m_\pi^2}{2s} \;.\nonumber
\ee
Changing the parity in the fermion sector can be done by adding a
$\gamma_5$ matrix. Then the basic operator
for the decay of a $1/2^+$ state into a nucleon and a pseudoscalar meson
 has the form:
\be
i\gamma_5 \hat k^\perp
\label{bas_12p}
\ee
where $\hat k^\perp$ is introduced just for convenience. Indeed
\be
\bar u(P) i\gamma_5\hat k^\perp u(k_1) &=&
\bar u(P) i\gamma_5(\hat k_1-a(s) \hat P) u(k_1)= \nonumber \\ &&
\bar u(P) i\gamma_5 u(k_1) (m_1+a(s) \sqrt s).
\label{last_chi}
\ee
Let us denote the last expression in (\ref{last_chi}) as $\chi$:
\be
\chi_i=m_i+a(s)\sqrt s \to {(\rm in\; c.m.s)}\; m_i+k_{i0} \;.
\ee
In general one can  also introduce another scalar expression using
$\gamma$ matrices and $k^\perp$:
\be
\varepsilon_{ijkl}\gamma_i\gamma_j k^\perp_k P_l
\ee
where $\varepsilon_{ijkl}$ is the
antisymmetric tensor. However using the
properties of the $\gamma$ matrices
\be
i\gamma^5 \gamma_i\gamma_j\gamma_k=\varepsilon_{ijkl}\gamma_l
\ee
one can show that this operator is identical to
(\ref{bas_12p}).

For the decay of systems with $J=L-1/2$
into $\pi N$ we obtain:
\be
\bar \Psi_{\mu_1\ldots\mu_{L-1}}
i\gamma_5 \gamma_\nu
X^{(L)}_{\nu\mu_1\ldots\mu_{L-1}}(k^\perp) u(k_1) \;.
\ee
Therefore the vertex function can be written as
\be
\bar \Psi_{\mu_1\ldots\mu_{L-1}}
N^-_{\mu_1\ldots\mu_{L-1}}(k^\perp) u(k_1) \;,\\
N^-_{\mu_1\ldots\mu_{L-1}}(k^\perp)=
i\gamma_5 \gamma_\nu
X^{(L)}_{\nu\mu_1\ldots\mu_{L-1}}(k^\perp) \;.
\ee
leading to the following amplitude for the transition $\pi N\to R\to
\pi N$:
\be
A&=&\bar u(k_1) X^{(L)}_{\alpha\mu_1\ldots\mu_{L-1}}(k)
\gamma^\perp_\alpha i\gamma_5
F^{\mu_1\ldots\mu_{L-1}}_{\nu_1\ldots\nu_{L-1}}(P)
i\gamma_5 \gamma^\perp_\xi \nonumber \\&&
X^{(L)}_{\xi\nu_1\ldots\nu_{L-1}}(q) u(q_1) BW_L^-(s) = \\ &&
(\sqrt{k_\perp^2}\sqrt{q_\perp^2})^{L-1} BW_L^-(s)
\bar u_i(k_1)\;\frac{\sqrt s+\hat P}{2\sqrt s}\;
\frac{\alpha(L)}{2L\!-\!1}
\nonumber \\ &&
\Big[ \hat k^\perp \hat q^\perp L\;P_{L-1}(z)-
\hat k^\perp\;
\frac{\sigma^\perp_{\mu\nu}k^\perp_\mu q^\perp_\nu}
{\sqrt{k_\perp^2}\sqrt{q_\perp^2}} \;\hat q^\perp
P'_{L-1}(z) \Big ] u_f(q_1)  \;. \nonumber
\ee
Taking into account that
\be
\hat k^\perp \hat q^\perp&=&(k^\perp q^\perp)+
\sigma^\perp_{\mu\nu} k^\perp_\mu q^\perp_\nu= \nonumber \\&&
\sqrt{k_\perp^2}\sqrt{q_\perp^2}
\left (z+\frac{\sigma^\perp_{\mu\nu} k^\perp_\mu q^\perp_\nu}
{\sqrt{k_\perp^2}\sqrt{q_\perp^2}}\right )
 \\
\hat k^\perp
\frac{\sigma^\perp_{\mu\nu} k^\perp_\mu q^\perp_\nu}
{\sqrt{k_\perp^2}\sqrt{q_\perp^2}}
\hat q^\perp &=&
\sqrt{k_\perp^2}\sqrt{q_\perp^2}
\left (1-z^2-z\frac{\sigma^\perp_{\mu\nu} k^\perp_\mu q^\perp_\nu}
{\sqrt{k_\perp^2}\sqrt{q_\perp^2}}\right ) \nonumber
\ee
(remember $z=(k^\perp q^\perp)/(\sqrt{k_\perp^2}\sqrt{q_\perp^2})$)
we obtain
\be
&A=&(\sqrt{k_\perp^2}\sqrt{q_\perp^2})^L BW_L^-(s) \bar
u_i(k_1)\;\frac{\sqrt s+\hat P}{2\sqrt s}\;
\frac{\alpha(L)}{L} \nonumber \\ &&\Big[
(LzP_{L-1}(z)-(1-z^2)P'_{L-1}(z))+ \nonumber \\
&&\frac{\sigma^\perp_{\mu\nu}k^\perp_\mu q^\perp_\nu}
{\sqrt{k_\perp^2}\sqrt{q_\perp^2}}(LP_{L-1}+zP'_{L-1}(z)) \Big ]
u_f(q_1).
\ee
Using the properties of Legendre polynomials (given in Appendix A)
the final expression for
$\pi N$ scattering due to '-' resonances reads:
\be
&A=&(\sqrt{k_\perp^2}\sqrt{q_\perp^2})^L BW_L^-(s) \bar
u_i(k_1)\;\frac{\sqrt s+\hat P}{2\sqrt s}
\frac{\alpha(L)}{L} \nonumber \\ &&\Big[ LP_L(z)+
\frac{\sigma^\perp_{\mu\nu}k^\perp_\mu q^\perp_\nu}
{\sqrt{k_\perp^2}\sqrt{q_\perp^2}}P'_{L}(z)\Big ]
u_f(q_1) \;.
\ee
Therefore the total $\pi N\to \pi N$ transition amplitude is equal to:
\be
&A=&(\sqrt{k_\perp^2}\sqrt{q_\perp^2})^L \bar
u_i(k_1)\!\frac{\sqrt s+\hat P}{2\sqrt s}
\left [ f_1-
\frac{\sigma^\perp_{\mu\nu}k^\perp_\mu q^\perp_\nu}
{\sqrt{k_\perp^2}\sqrt{q_\perp^2}}f_2 \right ]
u_f(q_1)
\nonumber \\
&f_1=&\sum\limits_L \Big [\frac{\alpha(L)}{2L\!+\!1} (L\!+\!1)BW_L^+(s)+
\frac{\alpha(L)}{L} L\; BW_L^-(s)\Big ] P_L(z)
\nonumber \\
&f_2=&\sum\limits_L \Big [\frac{\alpha(L)}{2L\!+\!1} BW_L^+(s)-
\frac{\alpha(L)}{L} BW_L^-(s)\Big ] P'_L(z)
\label{piN}
\ee
Let us calculate the amplitude (\ref{piN}) in the c.m.s. of the
resonance where $P=(\sqrt s,\vec 0)$:
\be
&&\bar u_i(k_1)\frac{\sqrt s\!+\!\hat P}{2\sqrt s}u_f(q_1)=
\nonumber \\
&&\frac{\left ( (k_{10}\!+\!m)\omega^*,-(\vec k_1\vec \sigma)\omega^*
\right )} {\sqrt{k_{10}\!+\!m}}
\left ( \begin{array}{cc} 1 & 0 \\ 0 & 0 \end{array} \right )
\frac{ \left ( \begin{array}{c}
(q_{10}\!+\!m)\omega' \\
(\vec q_1\vec \sigma)\omega'
\end{array}
\right )
}{\sqrt{q_{10}\!+\!m}}= \nonumber \\
 && \omega^*\sqrt{\chi_i\chi_f}\omega' \;,
\nonumber \\
&&\sigma_{\mu\nu}=
\left (
\begin{array}{cc}
-i\varepsilon_{\mu\nu j}\vec \sigma_j & 0 \\
0 & -i\varepsilon_{\mu\nu j}\vec \sigma_j
\end{array} \right )
=-i\varepsilon_{\mu\nu j}\vec \sigma_j\; I \;.
\ee
Here, $\omega$ and $\omega'$ are two-dimensional spinors of the initial
and final state nucleons. Thus
\be
A=(-1)^{L}(|\vec k||\vec q|)^L \sqrt{\chi_i\chi_f}\;
\omega^*\left [ f_1-i \varepsilon_{\mu\nu j}
\frac{\vec \sigma_j k_\mu q_\nu}{|\vec k||\vec q|} f_2 \right ]\omega'. \nonumber
\ee
Defining the vector normal to the decay plane as
\be
\vec n_j=-\varepsilon_{\mu\nu j}
\frac{k_\mu q_\nu}{|\vec k||\vec q|}
\ee
we obtain the final expression
\be
A=(-1)^{L}(|\vec k||\vec q|)^L \omega^*
\sqrt{\chi_i\chi_f}\;
\left [ f_1+i (\vec \sigma \vec n) f_2 \right]\omega \;.
\ee
When fitting $\pi N$ scattering data, the following expression
(defined in the c.m.s.) is often used
\be
&&A_{\pi N}=\omega^*\left [G(s,t)+H(s,t)i(\vec \sigma \vec n)
\right ]\omega' \;,
\nonumber \\
&&G(s,t)=\sum\limits_L \big [(L\!+\!1)F_L^+(s)- L F_L^-(s)\big ]
P_L(z) \;, \nonumber \\
&&H(s,t)=\sum\limits_L \big
[F_L^+(s)+ F_L^-(s)\big ] P'_L(z) \;.
\label{piN_others}
\ee
The $F^\pm_L$ are functions which depend only on energy.
Comparing our expressions with~(\ref{piN_others}) we obtain the following
correspondence:
\be
&F^+_L&=(-1)^{L+1}(|\vec k||\vec q|)^L
\sqrt{\chi_i\chi_f}\;
\frac{\alpha(L)}{2L\!+\!1} BW_L^+(s) \;,
\nonumber \\
&F^-_L&=(-1)^{L}(|\vec k||\vec q|)^L
\sqrt{\chi_i\chi_f}\;
\frac{\alpha(L)}{L}
BW_L^-(s)       \;.
\ee

\section{Operators for the decay of
baryons into a nucleon and a vector particle}

A vector particle (e.g. a virtual photon $\gamma^*$ or a $\rho$-meson)
has spin 1 and therefore the $\gamma^{*} N$ system
can form two spin states with $S=1/2$ and $3/2$.
In combination with the orbital
angular momentum, six sets of partial waves can be formed
\be
&J=L_{\gamma N}+\frac 12,\;  S=\frac 12,\; &P=(-1)^{L_{\gamma N}+1},
\; L_{\gamma N}=0,1,\ldots
\nonumber \\
&J=L_{\gamma N}-\frac 32,\;  S=\frac 32,\; &P=(-1)^{L_{\gamma N}+1},
\; L_{\gamma N}=2,3,\ldots
\nonumber \\
&J=L_{\gamma N}+\frac 12,\;  S= \frac 32,\; &P=(-1)^{L_{\gamma N}+1},
\; L_{\gamma N}=1,2,\ldots
\nonumber \\
&J=L_{\gamma N}- \frac 12,\;  S= \frac 12,\; &P=(-1)^{L_{\gamma N}+1},
\; L_{\gamma N}=1,2,\ldots
\nonumber \\
&J=L_{\gamma N}- \frac 12,\;  S= \frac 32,\; &P=(-1)^{L_{\gamma N}+1},
\; L_{\gamma N}=1,2,\ldots
\nonumber \\
&J=L_{\gamma N}+ \frac 32,\;  S= \frac 32,\; &P=(-1)^{L_{\gamma N}+1},
\; L_{\gamma N}=0,1,\ldots  \nonumber \\
\label{oper_set}
\ee

\subsection{Operators for \boldmath $1/2^-,\;3/2^+,\;5/2^-\ldots$ \unboldmath states}

Let us start from the operators for the '+' states.
A $1/2^-$ baryon decays into a baryon with $J^P=1/2^+$ and a
vector particle in either S or D-wave. In case of an S-wave
decay the orbital angular momentum operator is a unit operator
and the polarization vector
can be convoluted only with a $\gamma$ matrix. However the $\gamma$ matrix
changes the parity of the system. To compensate this
unwanted change an additional $\gamma_5$ matrix has to be
introduced. Therefore the operator describing
the transition of the state with spin $1/2^-$ into a $\gamma$ and $1/2^+$
fermion in S-wave is
\be
\bar u(P) i\gamma_\mu\gamma_5 u(k_1) \varepsilon_\mu  \;.
\label{ps_spin}
\ee
Here $\bar u(P)$ is the bispinor describing a baryon resonance
with momentum $P$,
$u(k_1)$  is the bispinor for the final fermion with momentum $k_1$
and $\varepsilon_\mu$ is the
polarization vector of the vector particle.
The operator (\ref{ps_spin}) is a
$1/2$ spin operator and its combination with the orbital angular momentum
operators $X^{(n)}_{\mu_1\ldots\mu_n}$ defines
the first set of the operators (\ref{oper_set}):
\be
\bar \Psi_{\alpha_1\ldots\alpha_L}
\gamma_\mu i\gamma_5
X^{(L)}_{\alpha_1\ldots\alpha_L}(k^\perp) u(k_1)  \varepsilon_\mu \;.
\label{ps_op1}
\ee
As before, $\Psi_{\alpha_1\ldots\alpha_L}$ is a fermionic bispinor
wave function with spin $J\!=\!L\!+\!1/2$,  and $k^\perp$
is the component of the
relative momentum of the
$\gamma^*N$ system orthogonal to the total
momentum of the system. For these partial waves the orbital angular momentum
in the $\gamma^* N$ system $L_{\gamma N}$ coincides with orbital angular momentum in
$\pi N$ which we denote as $L$. \\
The decay of a $1/2^-$ state into a $1/2^+$ and a vector particle in
D-wave must be described by the D-wave orbital angular
momentum operator:
\be
\bar u(P) \gamma_\nu i\gamma_5 X^{(2)}_{\mu\nu}(k^\perp)
u(k_1)  \varepsilon_\mu \;.
\ee
One can easily write down the whole set of such operators with
$J\!=\!L_{\gamma N}\!-\!3/2$ by
\be
\bar \Psi_{\alpha_1\ldots\alpha_{L}}
\gamma_\nu i \gamma_5
X^{(L+2)}_{\mu\nu\alpha_1\ldots\alpha_L}(k^\perp) u(k_1)\varepsilon_\mu \;.
\label{ps_op2}
\ee
Remember that $L$ is the orbital angular momentum in the decay
of a resonance into $\pi N$ ($L_{\gamma N}\!=\!L+2$). \\
The third set of operators starts from the total momentum $3/2$.
The basic operator describes the P-wave decay of a $3/2^+$ system
into a baryon and a vector particle.
It has the form
\be
\bar \Psi_{\mu} \gamma_\nu i\gamma_5 X^{(1)}_{\nu}(k^\perp) u(k_1)
\varepsilon_\mu   \;.
\ee
The operators for a baryon with $J\!=\!L_{\gamma N}\!+\!1/2$
can be written as
\be
\bar \Psi_{\mu\alpha_1\ldots\alpha_{L-1}}
\gamma_\nu i \gamma_5
X^{(L)}_{\nu\alpha_1\ldots\alpha_{L-1}}(k^\perp) u(k_1)  \varepsilon_\mu  \;.
\label{ps_op3}
\ee

In case of photoproduction rather than electroproduction
the operators (\ref{ps_op2}) are reduced due to
gauge invariance to those given in
(\ref{ps_op1}). Gauge invariance requires
\be
\varepsilon_\mu k_{1\mu}=
\varepsilon_\mu k_{2\mu}=
\varepsilon_\mu k^\perp_\mu=0  \;.
\ee
Using (\ref{fprop_2}) we obtain
\be
\bar \Psi_{\alpha_1\ldots\alpha_L}\gamma_\nu i \gamma_5
X^{(L+2)}_{\mu\nu\alpha_1\ldots\alpha_L}(k^\perp) u(k_1)
\varepsilon_\mu=   \;\;&&\\
\frac{-k^2_\perp\alpha(L)}{(2L\!-\!1)\alpha(L\!-\!2)}
\bar \Psi_{\alpha_1\ldots\alpha_L}\gamma_\mu i \gamma_5
X^{(L)}_{\alpha_1\ldots\alpha_L}(k^\perp) u(k_1) \varepsilon_\mu \;.&&
\nonumber
\ee
Although operators (\ref{ps_op2}) applied to the case of real photons
produce the same angular dependence as the
operators (\ref{ps_op1}), the former can provide an additional energy
dependence which can be important for broad states.

It is convenient
to write the decay amplitudes as a convolution of the bispinor wave
functions and the vertex functions
$V^{(i+)\mu}_{\alpha_1\ldots\alpha_L}$ $i=1,2,3$.
Then eqns.(\ref{ps_op1},\ref{ps_op2},\ref{ps_op3}) can be rewritten as
\be
\bar \Psi_{\alpha_1\ldots\alpha_L}
V^{(i+)\mu}_{\alpha_1\ldots\alpha_L}(k^\perp)
 u(k_1)  \varepsilon_\mu \;,  \nonumber
\ee
\be
&&V^{(1+)\mu}_{\alpha_1\ldots\alpha_L}(k^\perp)=
\gamma_\mu i\gamma_5
X^{(L)}_{\alpha_1\ldots\alpha_L}(k^\perp) \;,\\
&&V^{(2+)\mu}_{\alpha_1\ldots\alpha_L}(k^\perp)=
\gamma_\nu i \gamma_5
X^{(L+2)}_{\mu\nu\alpha_1\ldots\alpha_L}(k^\perp) \nonumber \;,\nn
&&V^{(3+)\mu}_{\alpha_1\ldots\alpha_L}(k^\perp)=
\gamma_\nu i \gamma_5
X^{(L)}_{\nu\alpha_1\ldots\alpha_{L-1}}(k^\perp)
g^\perp_{\mu\alpha_L} \;.\nonumber
\label{vf_plus}
\ee
In the helicity approach the property discussed above means that a
$1/2$ state is described by only one helicity amplitude while states
with higher spin are described by helicity $1/2$ and $3/2$ amplitudes.

\subsection{Operators for \boldmath $1/2^+,\;3/2^-,\;5/2^+\ldots$ \unboldmath states}

A $1/2^+$ particle decays into a fermion with $J^P=1/2^+$ and
spin-1 particle in relative P-wave only. The operator for spin $1/2$ of
the $\gamma^{*} N$ system can be constructed in the same way as the
corresponding operator for the '+'-states. The P-wave orbital
angular momentum operator must be convoluted with a $\gamma$-matrix. In
this case, the $\gamma_5$ operator is not needed to provide the correct
parity. The transition amplitude can be written as
\be
\bar u(P) \gamma_\xi \gamma_\mu X^{(1)}_\xi u(k_1) \varepsilon_\mu \;.
\label{ms_spin}
\ee
and the operator for the state with $S=1/2$ and
$J\;=L_{\gamma N}\;-\;1/2$ has the form
\be
\bar \Psi_{\alpha_1\ldots\alpha_{L-1}}
\gamma_\xi\gamma_\mu
X^{(L)}_{\xi\alpha_1\ldots\alpha_{L-1}}(k^\perp) u(k_1)
\varepsilon_\mu\; \;.\label{ms_op1}
\ee
with $L\equiv L_{\pi N}=L_{\gamma N}$. \\
For the 'minus' states,
the operators with $S=3/2$ and $J\!=\!L_{\gamma N}\!-\!1/2$
have the same
orbital angular momentum as the $S=1/2$ operator.
However here the polarization
vector convoluts with the
index of the orbital angular momentum operator. Then
\be
\bar
\Psi_{\alpha_1\ldots\alpha_{L-1}}
X^{(L)}_{\mu\alpha_1\ldots\alpha_{L-1}}(k^\perp) u(k_1)
\varepsilon_\mu \;.
\label{ms_op2}
\ee
The third set of operators starts from total spin
$3/2$. The basic operator describes the decay of the $3/2^-$ system
into the nucleon and a photon in relative S-wave. Thus
\be
\bar \Psi_{\mu} u(k_1) \varepsilon_\mu,
\ee
and we obtain the set:
\be
\bar \Psi_{\alpha_1\ldots\alpha_{L-1}}
X^{(L-2)}_{\alpha_2\ldots\alpha_{L-1}}(k^\perp)
g^\perp_{\alpha_1\mu}u(k_1)  \varepsilon_\mu
\label{ms_op3}
\ee
Remember that for these states $L=L_{\gamma N}+2$.

For real photons the operator (\ref{ms_op2}) vanishes for
$J=1/2^+$; for higher states these operators provide some
additional energy dependence in the
 partial waves (\ref{ms_op3}).
For convenience we introduce vertex functions
$V^{(i-)\mu}_{\alpha_1\ldots\alpha_{L-1}}$ $i=1,2,3$
as it was done in the case of '+' states
\be
\bar \Psi_{\alpha_1\ldots\alpha_{L-1}}
V^{(i-)\mu}_{\mu\alpha_1\ldots\alpha_{L-1}}(k^\perp)
 u(k_1)  \varepsilon_\mu \;,\nonumber
\ee
\be
&&V^{(1-)\mu}_{\alpha_1\ldots\alpha_{L-1}}(k^\perp)=
\gamma_\xi\gamma_\mu
X^{(L)}_{\xi\alpha_1\ldots\alpha_{L-1}}(k^\perp) \;,
\\
&&V^{(2-)\mu}_{\alpha_1\ldots\alpha_{L-1}}(k^\perp)=
X^{(L)}_{\mu\alpha_1\ldots\alpha_{L-1}}(k^\perp) \;,
\nonumber \\
&&V^{(3-)\mu}_{\alpha_1\ldots\alpha_{L-1}}(k^\perp)=
X^{(L-2)}_{\alpha_2\ldots\alpha_{L-1}}(k^\perp)
g^\perp_{\alpha_1\mu} \;.\nonumber
\label{vf_minus}
\ee

\section{Single meson photoproduction}

The amplitude for the photoproduction of a single pseudoscalar meson
(for the sake of simplicity let us take the pion) is
well known and can be found in the literature (see for example
\cite{tyator} and references therein). The general structure of the
amplitude is
$$A=\omega^*J_\mu\varepsilon_\mu \omega' \;,$$
\be
\label{mult_1}
&&
J_\mu = \\&&
i {\mathcal F_1}
 \sigma_\mu +{\mathcal F_2} (\vec \sigma \vec q)
\frac{\varepsilon_{\mu i j} \sigma_i k_j}{|\vec k| |\vec q|}
+i {\mathcal F_3} \frac{(\vec \sigma \vec k)}{|\vec k| |\vec q|} q_\mu
+i {\mathcal F_4} \frac{(\vec \sigma \vec q)}{\vec q^2} q_\mu  \;.\nonumber
\ee
where $\vec q$ is the momentum of the nucleon in the $\pi N$ channel
and $\vec k$ is the
momentum of the nucleon in the $\gamma N$ channel calculated in  the c.m.s. of
the reaction and $\sigma_i$ are Pauli matrices.

The functions ${\mathcal F_i}$ have the following angular dependence:
\be
&&{\mathcal F_1}(z) = \nonumber \\&&
 \sum^{\infty}_{L=0} [LM^+_L+E^+_L]
P^{\prime}_{L+1}(z) +
[(L+1)M^-_L + E^-_L]
P^{\prime}_{L-1}(z) \;, \nonumber \\&&
{\mathcal F_2}(z) = \sum^{\infty}_{L=1} [(L+1)M^+_L+LM^-_L]
P^{\prime}_{L}(z)   \;,\nonumber  \\&&
{\mathcal F_3}(z) = \sum^{\infty}_{L=1} [E^+_L-M^+_L]
P^{\prime\prime}_{L+1}(z) + [E^-_L + M^-_L]
P^{\prime\prime}_{L-1}(z)\;, \nonumber \\&&
{\mathcal F_4} (z) = \sum^{\infty}_{L=2} [M^+_L - E^+_L - M^-_L
-E^-_L] P^{\prime\prime}_{L}(z).
\label{mult_2}
\ee
Here $L$ corresponds to the orbital angular momentum in the $\pi N$ system,
$P_L(z)$ are Legendre polynomials
$z=(\vec k\vec q)/(|\vec  k||\vec q|)$ and $E^\pm_L$ and $M^\pm_L$ are
electric and magnetic multipoles describing transitions to
states with $J=L\pm 1/2$. There are no contributions from $M^+_0$,
$E^-_0$ and $E^-_1$ for spin 1/2 resonances. In the following we will
construct the $\gamma N\to \pi N$ transition amplitudes using the
operators defined in the previous sections and show that in c.m.s.
these amplitudes satisfy the equations (\ref{mult_1}, \ref{mult_2}).

\subsection{Photoproduction amplitudes for \boldmath
$1/2^-,\;3/2^+,\;5/2^-\ldots$ \unboldmath states}

The angular dependence of the single-meson-production amplitude via
an intermediate resonance has the general form
\be
\bar u(q_1)
\tilde N^\pm_{\alpha_1\ldots\alpha_n}(q^\perp)
F^{\alpha_1\ldots\alpha_n}_{\beta_1\ldots\beta_n}(P)
V^{(i\pm)\mu}_{\beta_1\ldots\beta_n}(k^\perp)
u(k_1)  \varepsilon_\mu \;.
\ee
Here $q_1$ and $k_1$ are the momenta of the nucleon in the $\pi N$
and $\gamma N$ channel and $q^\perp$ and $k^\perp$ are
the components of the relative momenta which are orthogonal to
the total momentum of the resonance.

If states with $J=L+1/2$ are produced from a $\gamma N$ partial wave
with spin $1/2$ one has the following expression for
the amplitude:
\be
&& A^+(1/2)=  \bar u(q_1)
X^{(L)}_{\alpha_1\ldots\alpha_L}(q^\perp) \nonumber\\ &&
F^{\alpha_1\ldots\alpha_L}_{\beta_1\ldots\beta_L}(P)
\gamma_\mu i \gamma_5
X^{(L)}_{\beta_1\ldots\beta_L}(k^\perp)
u(k_1)  \varepsilon_\mu
BW(s) \;.
\label{am_ps_op1}
\ee
where $BW(s)$ represent  the dynamical part of the amplitude.
Taking into account the properties of the projection operator this
expression can be rewritten as
\be
&&\bar u(q_1)
X^{(L)}_{\alpha_1\ldots\alpha_L}(q^\perp)
T^{\alpha_1\ldots\alpha_L}_{\beta_1\ldots\beta_L}
\frac{\sqrt s+\hat P}{2\sqrt s}
X^{(L)}_{\beta_1\ldots\beta_L}(k^\perp)
\nonumber \\
&&\gamma_\mu i \gamma_5 u(k_1)  \varepsilon_\mu =
\nonumber \\
&&\bar u(q_1)\Big [
\frac{L\!+\!1}{2L\!+\!1}
X^{(L)}_{\alpha_1\ldots\alpha_L}(q^\perp)
X^{(L)}_{\alpha_1\ldots\alpha_L}(k^\perp)-\nonumber \\
&&\frac{L}{2L\!+\!1} \sigma_{\alpha\beta}
X^{(L)}_{\alpha\alpha_2\ldots\alpha_L}(q^\perp)
X^{(L)}_{\beta\alpha_2\ldots\alpha_L}(k^\perp)\Big ]
\nonumber \\
&&\frac{\sqrt s+\hat P}{2\sqrt s}
\gamma_\mu i \gamma_5 u(k_1)  \varepsilon_\mu \;.
\ee
Using the expression for the convolution of two
$X$-operators with two
external indices (as given in Appendix B) one obtains
\be
A^+(1/2)=\bar u(q_1)
\frac{L\!+\!1}{2L\!+\!1}\alpha(L)
(\sqrt{q^\perp}\sqrt{k^\perp})^L
\nonumber \\
\Big [ P_L(z)-\frac{P'_L(z)}{L\!+\!1} \sigma_{\alpha\beta}
\frac{q_\alpha^\perp k_\beta^\perp}
{(\sqrt{q^\perp}\sqrt{k^\perp})} \Big ]
\nonumber \\
\frac{\sqrt s+\hat P}{2\sqrt s}
\gamma_\mu i \gamma_5 u(k_1)  \varepsilon_\mu
BW(s) \;.
\ee
In the c.m.s.
\be
\bar u(q_1)
\frac{\sqrt s+\hat P}{2\sqrt s}
\gamma_\mu i \gamma_5 u(k_1) \varepsilon_\mu =
-\sqrt{\chi_i\chi_f}i\omega^*(\vec\varepsilon_i\vec \sigma_i)\omega' \;.
\ee
holds, leading to
\be
&& A^+(1/2)=\omega^*\sqrt{\chi_i\chi_f}\;
\frac{L\!+\!1}{2L\!+\!1}\alpha(L) i (-\varepsilon_i)
(\sqrt{q^\perp}\sqrt{k^\perp})^L
\nonumber \\ &&
\Big [\sigma_i
P_L(z)+
i\frac{P'_L(z)}{L\!+\!1} \varepsilon_{\alpha\beta\xi}
\sigma_\xi \sigma_i
\frac{q_\alpha^\perp k_\beta^\perp}
{(\sqrt{q^\perp}\sqrt{k^\perp})} \Big ]\omega'
BW(s).
\ee
Here all vectors are three-dimensional.
Using in addition the properties of Pauli matrices
\be
\sigma_i\sigma_j=\delta_{ij}+i\varepsilon_{ijk}\sigma_k
\ee
one obtains the final expression
\be
&& A^+(1/2)=-\omega^*\sqrt{\chi_i\chi_f}\;
\frac{\alpha(L)}{2L\!+\!1} \varepsilon_i
(\sqrt{q^\perp}\sqrt{k^\perp})^L
\nonumber \\ &&
\Big [i\sigma_i\Big ( (L\!+\!1)P_L(z)+zP'_L(z) \Big ) +
(\vec \sigma\vec q)\frac{\varepsilon_{ijm}\sigma_j k_m}
{|\vec k||\vec q|}P'_L(z) \Big ]
\nonumber \\ &&
\omega'
BW(s)\;.
\ee
 Taking into account the properties of the Legendre polynomials
(given in Appendix A)
the amplitude can be compared with equations (\ref{mult_1}),
(\ref{mult_2}). One finds the following correspondence between the spin
operators and multipoles:
\be
E^{+(\frac12)}_L&=&(-1)^L \sqrt{\chi_i\chi_f}\;
\frac{\alpha(L)}{2L\!+\!1}
\frac{(|\vec k||\vec q|)^L}{L\!+\!1}
BW(s) \;,
\nonumber \\
M^{+(\frac12)}_L&=&E^{+(\frac12)}_L \;.
\label{mpd_p1}
\ee
Here and below $E^{+(\frac12)}_L$ and $M^{+(\frac12)}_L$ multipoles correspond
to the decomposition of spin 1/2 amplitudes.
In the case of photoproduction, only two $\gamma N$
operators are independent for every resonance with spin $3/2$ and higher
(for $J=1/2$ states there is only one independent operator). For the set of
$J=L+1/2$ states the second operator has the amplitude
structure
\be
&& A^+(3/2)=
\bar u(q_1)
X^{(L)}_{\alpha_1\ldots\alpha_L}(q^\perp)
F^{\alpha_1\ldots\alpha_L}_{\mu\beta_2\ldots\beta_L}(P)
\nonumber \\ &&
\gamma_\xi i \gamma_5
X^{(L)}_{\xi\beta_2\ldots\beta_L}(k^\perp)
u(k_1)  \varepsilon_\mu
BW(s)   \;.
\label{mult_p_op2}
\ee
Using expressions given in Appendix B one obtains the
multipole decomposition
\be
E^{+(\frac32)}_L&=&(-1)^L
\sqrt{\chi_i\chi_f}\;
\frac{\alpha(L)}{2L\!+\!1}
\frac{(|\vec k||\vec q|)^L}{L\!+\!1}
BW(s) \;,
\nonumber \\
M^{+(\frac32)}_L&=&-\frac{E^{+(\frac32)}_L}{L} \;.
\label{mpd_p2}
\ee
Here and below $E^{+(\frac32)}_L$ and $M^{+(\frac32)}_L$ multipoles correspond
to the decomposition of spin 3/2 amplitudes.

\subsection{Photoproduction amplitudes for  \boldmath
$1/2^+,\;3/2^-,\;5/2^+\ldots$ \unboldmath states}

The $\gamma N\to \pi N$ amplitude for
states with $J=L-1/2$ in the $\pi N$ channel has the
structure
\be
&& A^-(1/2)=\bar u(q_1) \gamma_\xi i\gamma_5
X^{(L)}_{\xi\alpha_1\ldots\alpha_{L-1}}(q^\perp)
F^{\alpha_1\ldots\alpha_{L-1}}_{\beta_1\ldots\beta_{L-1}}(P)
\nonumber \\ &&
\gamma_\xi \gamma_\mu
X^{(L)}_{\xi\beta_1\ldots\beta_{L-1}}(k^\perp)
u(k_1)  \varepsilon_\mu BW(s)  \;.
\label{mult_m_op1}
\ee
For amplitude (\ref{mult_m_op1}) we find the following
correspondence to the multipole decomposition
(see Appendix B for details):
\be
E_L^{-(\frac12)}&=&(-1)^{L}
\sqrt{\chi_i\chi_f}\;
|\vec k|^L |\vec q|^L\frac{\alpha(L)}{L^2} BW(s)\;,
\nn
M_L^{-(\frac12)}&=&-E_L^{-(\frac12)}\;.
\label{mpd_n1}
\ee
Amplitudes including spin $3/2$ operators have the structure
\be
&& A^-(3/2)=\bar u(q_1) \gamma_\xi i \gamma_5
X^{(L)}_{\xi\alpha_1\ldots\alpha_{L-1}}(q^\perp)
F^{\alpha_1\ldots\alpha_{L-1}}_{\mu\beta_2\ldots\beta_{L-1}}(P)
\nonumber \\ &&
X^{(L-2)}_{\beta_2\ldots\beta_{L-1}}(k^\perp)
u(k_1)  \varepsilon_\mu BW(s) \;.
\label{mult_m_op2}
\ee
Using expressions in Appendix B the decomposition of this amplitude into the multipole
representation is the following:
\be
E_L^{-(\frac32)}&=&(-1)^{L}
\sqrt{\chi_i\chi_f}\;
|\vec k|^{L-2} |\vec q|^L\frac{\alpha(L-2)}{(L\!-\!1)L}
BW(s) \;,
\nn
M_L^{-(\frac32)}&=&0  \;.
\label{mpd_n2}
\ee

\subsection{Relations between the amplitudes in the spin-orbit
and helicity representation}

The helicity transition amplitudes
are combinations of
the spin 1/2 and 3/2 amplitudes $A^\pm(1/2)$, $A^\pm(3/2)$. For '+'
multipoles the relations between the helicity amplitudes and multipoles
are
\be
\tilde A^{1/2}&=-&\frac 12 \big (L M^+_L+(L+2)E_L^+\big )\;,
\nn
\tilde A^{3/2}&= &\frac 12 \sqrt{L(L\!+\!2)}\big (E^+_L-M_L^+ \big) \;.
\ee
For the '-' sector the relations are
\be
\tilde A^{1/2}&= &\frac 12 \big ((L+1)M^-_L-(L-1)E_L^- \big ) \;,
\nn
\tilde A^{3/2}&=-&\frac 12 \sqrt{(L\!-\!1)(L\!+\!1)}\big (E^-_L+M_L^-) \;.
\ee
The energy dependence of the helicity transition amplitudes
$\tilde A^{1/2}$ and $\tilde A^{3/2}$ is a model dependent subject
which will be discussed in our forthcoming paper.
In the mass of a resonance these amplitudes are connected with
helicity vertex functions $A^{1/2},A^{3/2}$ given in PDG
by a constant:
\be
(A^{1/2},A^{3/2})=C(\tilde A^{1/2},\tilde A^{3/2})
\ee
which (together with resonance parameterization) can be found
for example in \cite{hel_factor}.
The ratio of the transition amplitudes
$\tilde A^{1/2},\tilde A^{3/2}$
(which is equal to the ratio of the helicity vertex
functions in the case of the Breit-Wigner parameterization) depends on
the $\gamma$-nucleon interaction only and should be the same in all
photoproduction reactions.

For '+' states we obtain the following decomposition of the spin 1/2
amplitude (\ref{mpd_p1}):
\be
\tilde A^{1/2}&=-&(L+1) E^{+(\frac12)}_L \;,
\nn
\tilde A^{3/2}&= &0  \;.
\label{ratio_01}
\ee
Obviously the spin $1/2$ state can not have a
helicity $3/2$ projection.
For the spin $3/2$ state one gets
\be
\tilde A^{1/2}&=-&\frac{L+1}{2}E^{+(\frac32)}_L  \;,
\nn
\tilde A^{3/2}&= &\frac 12 \sqrt{\frac{L+2}{L}}(L+1)E^{+(\frac32)}_L \;.
\label{ratio_02}
\ee

The ratio of the helicity amplitudes can be calculated directly if the
ratio of the spin amplitudes is known.
The $BW(s)$ in both amplitudes
is an energy dependent part of the amplitude which
depends on the model used in the analysis.
If a resonance is produced and decays with radius $r$ the
regularization of the amplitude can be done with, e.g.,
Blatt-Weisskopf formfactors (see Appendix C).
If we also explicitly extract the initial coupling constants
$g_{1/2}$ and $g_{3/2}$ for the spin 1/2 and 3/2,
then the expression for the total amplitude for '+' states has the form
\be
&& A^{L+}_{tot}= \\ && \big [ g_{1/2}\;A^+(1/2)+
g_{3/2}A^+(3/2)\big ] \frac{1}
{F(L,q^2_\perp,r) F(L,k^2_\perp,r)}.  \nonumber
\ee
In this case the  multipole amplitudes can be rewritten as following:
\be
&& E^{+(\frac12)}_L=
 \\ &&
(-1)^L \sqrt{\chi_i\chi_f}\;
\frac{\alpha(L)}{2L\!+\!1}
\frac{(|\vec k||\vec q|)^L}{L\!+\!1}
\frac{g_{1/2} BW(s)}
{F(L,q^2_\perp,r) F(L,k^2_\perp,r)} \;,
 \nonumber
\label{mpd_p1a}
\ee
\be
&& E^{+(\frac32)}_L =
 \\ &&
(-1)^L
\sqrt{\chi_i\chi_f}\;
\frac{\alpha(L)}{2L\!+\!1}
\frac{(|\vec k||\vec q|)^L}{L\!+\!1}
\frac{g_{3/2} BW(s)}
{F(L,q^2_\perp,r) F(L,k^2_\perp,r)}  \;,
 \nonumber
\label{mpd_p2a}
\ee
\be
E^{+}_L =  E^{+(\frac12)}_L + E^{+(\frac32)}_L \;.
\ee
From (\ref{ratio_01}) and (\ref{ratio_02}) one can calculate the
the ratio between helicity amplitudes for '+' states:
\be
\frac{\tilde A^{3/2}}{\tilde A^{1/2}} =
\frac{A^{3/2}}{A^{1/2}}=
 -\;\frac{\frac 12 \sqrt{\frac{L+2}{L}}(L+1)E^{+(\frac32)}_L}
{\frac{L+1}{2}E^{+(\frac32)}_L\; +\; (L+1) E^{+(\frac12)}_L} =
 \nonumber \\
-\sqrt{\frac{L+2}{L}}\;\frac{1}{1+2R}\;\;,
\qquad R=\frac{g_{1/2}}{g_{3/2}}\quad
\ee
 This ratio does not depend on the final state of the photoproduction
 process, is valid for any photoproduction reaction and should be compared
with PDG values.

 In the case of the '-' states we get for the spin $1/2$ amplitude:
\be
\tilde A^{1/2}&=-& L E^{-(\frac12)}_L  \;,
\nn
\tilde A^{3/2}&= &0 \;.
\ee
and for the spin $3/2$ amplitudes
\be
\tilde A^{1/2}&=-&\frac{L-1}{2}E^{-(\frac32)}_L \;,
\nn
\tilde A^{3/2}&=-&\frac 12 \sqrt{(L-1)(L+1)}E^{-(\frac32)}_L \;.
\ee
For $(-)$ states the $\gamma p$ vertex has the same orbital momentum
as the $\pi N$ vertex for spin $\frac 12$ amplitudes, and $L-2$
for spin $\frac 32$ amplitudes
 \be
A^{L-}_{tot}=\Big [ \frac{g_{1/2}A^-(1/2)}{F(L,k^2_\perp,r)}+
\frac{g_{3/2}A^+(3/2)}{F(L\!-\!2,k^2_\perp,r)}
 \Big ] \frac{1}{F(L,q^2_\perp,r)}  \;.\nonumber
 \ee
The  multipole amplitudes can be rewritten as follows:
\be
&& E_L^{-(\frac12)} = \\ && (-1)^{L}
\sqrt{\chi_i\chi_f}\;
|\vec k|^L |\vec q|^L\frac{\alpha(L)}{L^2}\;
\frac{g_{1/2}BW(s)}{F(L,q^2_\perp,r)\;F(L,k^2_\perp,r)}  \;,\nonumber
\label{mpd_n1a}
\ee
\be
&& E_L^{-(\frac32)}=(-1)^{L}
\sqrt{\chi_i\chi_f}\;
|\vec k|^{L-2} |\vec q|^L\frac{\alpha(L-2)}{(L\!-\!1)L}\; \nonumber \\ &&
\frac{g_{3/2}BW(s)}{F(L,q^2_\perp,r)\;F(L-2,k^2_\perp,r)} \;,
\label{mpd_n2a}
\ee
\be
E^{-}_L =  E^{-(\frac12)}_L + E^{-(\frac32)}_L \;.
\ee
For the ratio of helicity amplitudes one obtains:
\be
\frac{\tilde A^{3/2}}{\tilde A^{1/2}} =
\frac{A^{3/2}}{A^{1/2}} =
\frac{\frac 12 \sqrt{(L-1)(L+1)}E^{-(\frac32)}_L}
{\frac{L-1}{2}E^{-(\frac32)}_L \;+\; L E^{-(\frac12)}_L} =
\nonumber \\
\sqrt{\frac{L+1}{L-1}}\;
\frac{1}{1+2R\;\kappa}  \;,
\ee
where
\be
\kappa=\frac{(2L-1)(2L-3)}{L(L-1)}|\vec k|^2
\frac{F(L\!-\!2,k^2_\perp,r)}{F(L,k^2_\perp,r)}
\ee
This ratio calculated in the resonance mass should be compared
with PDG values.

\subsection{Operators for \boldmath $1/2^-,\;3/2^+,\;5/2^-\ldots$ \unboldmath states}

A $1/2^-$ particle decays into a $J^P=3/2^+$-particle and
pseudoscalar meson  in  D-wave.
Only one of the indices of the orbital angular momentum operator
can  be absorbed by  a
$\gamma$-matrix. Again, to compensate the change of parity
due to the $\gamma$-matrix one has to introduce an additional
$\gamma_5$-matrix. The operator describing
the transition of a state with spin $1/2^-$ into a $0^-$ and a $3/2^+$
state is
\be
\bar u(P)\; i\gamma_5 \gamma_\nu
X^{(2)}_{\mu\nu} \Psi^{\Delta}_{\mu} \;.
\label{ps_spin1}
\ee
where $\bar u_p$ is a bispinor describing an initial state and
$\Psi^{\Delta}_{\mu}$  is a vector bispinor for the final spin-3/2 fermion.
The first set of operators
derived from eq.(\ref{ps_spin1}) reads
\be
\bar \Psi_{\alpha_1\ldots\alpha_{L_\Delta-2}}\;i\gamma_5 \gamma_\nu
X^{(L_\Delta)}_{\mu\nu\alpha_1\ldots\alpha_{L_\Delta-2}}
\Psi^{\Delta}_{\mu}\;,
\;\; L_\Delta=2,3,\ldots
\ee
However it is again convenient to rewrite this expression using the
orbital angular momentum $L$. In this case $L_\Delta\!=\!L\!+2$, and
\be
\bar \Psi_{\alpha_1\ldots\alpha_L}\;
i\gamma_5 \gamma_\nu
X^{(L+2)}_{\mu\nu\alpha_1\ldots\alpha_{L}}
\Psi^{\Delta}_{\mu} \;,
\qquad  L=0,1,\ldots
\label{ps_op11}
\ee
The second set of operators starts from total spin $3/2$.
The basic operator describes the decay of the $3/2^+$ system
into $\Delta$ and pion in a P-wave. It has the
form
\be
\bar \Psi_{\alpha}\;i \gamma_5 \gamma_\nu X^{(1)}_{\nu}
g^\perp_{\alpha\mu}\Psi^{\Delta}_{\mu} \;.
\ee
The second set of the operators
can be written as (here $L_\Delta=L$)
\be
\bar \Psi_{\alpha_1\ldots\alpha_L}\;i
\gamma_5\gamma_\nu X^{(L)}_{\nu \alpha_2\ldots\alpha_L}
g^\perp_{\alpha_1\mu}
\Psi^{\Delta}_{\mu} \;,
\qquad  L=1,2,\ldots
\label{ps_op31}
\ee

Thus the vertex functions for '+' states are
\be
\bar \Psi_{\alpha_1\ldots\alpha_L}
N^{(i+)\mu}_{\alpha_1\ldots\alpha_L}
\Psi^{\Delta}_{\mu} \;,
\qquad
&&N^{(1+)\mu}_{\alpha_1\ldots\alpha_L}=
i\gamma_5 \gamma_\nu
X^{(L+2)}_{\mu\nu\alpha_1\ldots\alpha_{L}}   \;,
\nonumber \\
&&N^{(2+)\mu}_{\alpha_1\ldots\alpha_L}=
i\gamma_5 \gamma_\nu
X^{(L)}_{\nu \alpha_2\ldots\alpha_L}
g^\perp_{\alpha_1\mu}  \;.
\nonumber \\
\label{vf_pD_plus}
\ee

\subsection{Operators for \boldmath $1/2^+,\;3/2^-,\;5/2^+\ldots$ \unboldmath states}

A $1/2^+$ particle may decay into a $J^P=3/2^+$ baryon and $0^-$
meson in P-wave. In this case the P-wave orbital angular
momentum operator must be converted with the vector bispinor
$\Psi^{\Delta}_{\mu}$.  The $\gamma_5$ operator is not needed to provide a correct parity for the
state. Then
\be
\bar u(P) X^{(1)}_\mu \Psi^{\Delta}_{\mu}\;.
\label{ms_spin1}
\ee
The operator for the state with $S=3/2$ and $J=L-1/2$ ($L=L_\Delta$)
has the form
\be
\bar \Psi_{\alpha_1\ldots\alpha_{L-1}}
X^{(L)}_{\mu\alpha_1\ldots\alpha_{L-1}}
\Psi^{\Delta}_{\mu} \;,
\qquad \qquad L=1,2,\ldots
\label{ms_op11}
\ee
As before, the second set of operators starts from total spin
$S=3/2$. The basic operator describes the decay of the $3/2^-$ system
into a $3/2^+$ particle and pion in S-wave. Thus
\be
\bar \Psi_{\mu} \Psi^{\Delta}_{\mu},
\ee
and we obtain for this set
\be
\bar \Psi_{\mu\alpha_1\ldots\alpha_{L_\Delta}}
X^{(L_\Delta)}_{\alpha_1\ldots\alpha_{L_\Delta}} \Psi^{\Delta}_{\mu}   \;,
\qquad  \qquad L_{\Delta}=0,1,\ldots
\ee
Here $L=L_\Delta+2$ and the amplitude can be rewritten as
\be
\bar \Psi_{\alpha_1\ldots\alpha_{L-1}}
X^{(L-2)}_{\alpha_2\ldots\alpha_{L-1}} g^\perp_{\alpha_1 \mu}
\Psi^{\Delta}_{\mu} \;,
\qquad L=2,3,\ldots
\label{ms_op31}
\ee
The vertex functions for '-' states are given by:
\be
\bar \Psi_{\alpha_1\ldots\alpha_{L-1}}
N^{(i-)\mu}_{\alpha_1\ldots\alpha_{L-1}}
\Psi^{\Delta}_{\mu} \;,
\qquad
&&N^{(1-)\mu}_{\alpha_1\ldots\alpha_{L-1}}=
X^{(L)}_{\mu\alpha_1\ldots\alpha_{L-1}} \;,
\nonumber \\
&&N^{(2-)\mu}_{\alpha_1\ldots\alpha_{L-1}}=
X^{(L-2)}_{\alpha_2\ldots\alpha_{L-1}} g^\perp_{\alpha_1 \mu} \;.
\nonumber \\
\label{vf_pD_minus}
\ee

\subsection{Operators for the decay into states with different parity}

The operators given in the previous sections provide a full set of
operators for the decay of a baryon into meson with spin 0 and fermion
with spin $3/2$. Indeed, for construction of operators only  the
total spin of the system plays the role. Thus the operators for
$J^+\to 0^-+3/2^+$ decays have the same form as the operators for
$J^+\to 0^++3/2^-$, $J^-\to 0^++3/2^+$ and $J^-\to 0^-+3/2^-$ decays.

\section{Double pion photoproduction amplitudes }

Let us construct the amplitudes for double pion photoproduction. Here
reactions as shown in Fig.~\ref{double} are taken into account where
the decay into the final state proceeds via production an intermediate
baryon or meson resonance.
\begin{figure}[hb]
\centerline{\epsfig{file=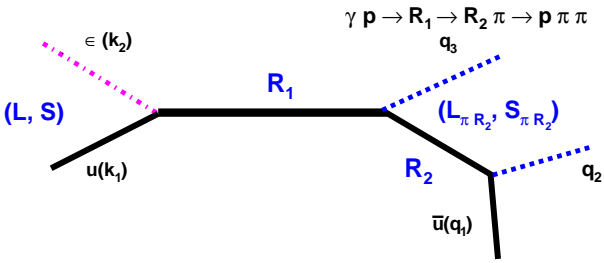,width=12.cm,clip=on}}
\caption{Photoproduction of two mesons due to the cascade
of a resonance}
\label{double}
\end{figure}
The general form of the angular dependent part of the amplitude for
such a process is
\be
&&\bar u(q_1)
 \tilde N_{\alpha_1\ldots\alpha_n}(R_{2}\!\to\!\mu N)
F^{\alpha_1\ldots\alpha_n}_{\beta_1\ldots\beta_n}(q_1+q_2)
\nonumber \\
&&\tilde N^{(j)\beta_1\ldots\beta_n}_{\gamma_1\ldots\gamma_m}
(R_1\!\to\!\mu R_{2})
F^{\gamma_1\ldots\gamma_m}_{\xi_1\ldots\xi_m}(P)
V^{(i)\mu}_{\xi_1\ldots\xi_m}(R_1\!\to\!\gamma N)
\nonumber\\ &&
u(k_1)  \varepsilon_\mu, \qquad\qquad P=q_1+q_2+q_3=k_1+k_2
\label{double1}
\ee
The resonance $R_1$ with spin $J=m+1/2$ is produced in $\gamma N$
interaction, propagates and then decays into a meson  and a baryon
resonance $R_2$ with spin $J=n+1/2$. Then the resonance  $R_2$
propagates and decays into the final meson and a nucleon.  %

In the following the full vertex functions used for the
construction of amplitudes are given here for convenience
of the reader. One should remember that
the $\tilde N$ functions are different from $N$-functions by the order
of $\gamma$-matrices. For $R\to 0^-\!+1/2^+$ transitions
\be
\tilde N^+_{\mu_1\ldots\mu_n}=X^{(n)}_{\mu_1\ldots\mu_n} \;\; \tilde
N^-_{\mu_1\ldots\mu_n}=i\gamma_\nu \gamma_5
X^{(n+1)}_{\nu\mu_1\ldots\mu_n}
\label{sum_v1}
\ee
holds, while we have
\be
\begin{array}{ll}
\tilde N^{(1+)\mu}_{\alpha_1\ldots\alpha_n}=
i\gamma_\nu \gamma_5
X^{(n+2)}_{\mu\nu\alpha_1\ldots\alpha_{n}} \;,\qquad
&\tilde N^{(1-)\mu}_{\alpha_1\ldots\alpha_{n}}=
X^{(n+1)}_{\mu\alpha_1\ldots\alpha_{n}} \;,\\
\tilde N^{(2+)\mu}_{\alpha_1\ldots\alpha_n}=
i\gamma_\nu \gamma_5
X^{(n)}_{\nu \alpha_2\ldots\alpha_n}
g^\perp_{\alpha_1\mu}\;,\qquad
&\tilde N^{(2-)\mu}_{\alpha_1\ldots\alpha_{n}}=
X^{(n-1)}_{\alpha_2\ldots\alpha_{n}} g^\perp_{\alpha_1 \mu}\;.
\end{array}
\nonumber\\
\label{sum_v2}
\ee
for $R\to 0^-\!+3/2^+$ transitions, and
\be
\begin{array}{ll}
V^{(1+)\mu}_{\alpha_1\ldots\alpha_n}=
\gamma_\mu i\gamma_5
X^{(n)}_{\alpha_1\ldots\alpha_n} \;,\qquad &
V^{(1-)\mu}_{\alpha_1\ldots\alpha_{n}}=
\gamma_\xi\gamma_\mu
X^{(n+1)}_{\xi\alpha_1\ldots\alpha_{n}} \;,
\\
V^{(2+)\mu}_{\alpha_1\ldots\alpha_n}=
\gamma_\nu i \gamma_5
X^{(n+2)}_{\mu\nu\alpha_1\ldots\alpha_n} \;,\qquad &
V^{(2-)\mu}_{\alpha_1\ldots\alpha_{n}}=
X^{(n+1)}_{\mu\alpha_1\ldots\alpha_{n}}  \;,
\\
V^{(3+)\mu}_{\alpha_1\ldots\alpha_n}=
\gamma_\nu i \gamma_5
X^{(n+1)}_{\nu\alpha_1\ldots\alpha_{n}}
g^\perp_{\mu\alpha_n}\;,\qquad &
V^{(3-)\mu}_{\alpha_1\ldots\alpha_{n}}=
X^{(n-1)}_{\alpha_2\ldots\alpha_{n}}
g^\perp_{\alpha_1\mu}   \;.
\end{array}
\nonumber\\
\label{sum_v3}
\ee
for $R\to 1^-\!+1/2^+$ transitions. Here $n$ is related to the total
spin of the resonance by $J=n+1/2$.

\subsection{Amplitudes for baryons states decaying into a \boldmath
$1/2$ \unboldmath state and a pion}

In this section explicit expressions for the angular dependent
part of the amplitudes are given for the case of a baryon produced
in a $\gamma^* N$ collision. The baryon decays into a pseudoscalar
particle and another (intermediate) baryon with spin 1/2 (decaying in
turn into meson and nucleon).

\subsubsection{The $1/2^-,\;3/2^+,\;5/2^-\ldots$  states}

The amplitude for a '+' state ($R_1$) produced in a $\gamma^*N$
collision in a partial wave $(i)$ decaying
into a $0^-$ meson and an intermediate $1/2^+$ baryon ($R_2$) has the form
\be
&&A^{(i)}=
\bar u(q_1)\tilde N^-(q^\perp_{12})\;
\frac{\hat q_1\! +\! \hat q_2\! +\! \sqrt {s_{12}}}{2\sqrt{s_{12}}}
\tilde N^+_{\alpha_1\ldots\alpha_L}(q^\perp_1)
\nn &&
F^{\alpha_1\ldots\alpha_L}_{\beta_1\ldots\beta_L}(P)
V^{(i+)\mu}_{\beta_1\ldots\beta_L}(k^\perp)
u(k_1)  \varepsilon_\mu
\nn &&
=\bar u(q_1)\;i\hat q^\perp_{12}\gamma_5  \;
\frac{\hat q_1\! +\! \hat q_2\! +\! \sqrt {s_{12}}}{2\sqrt{s_{12}}}\;
X^{(L)}_{\alpha_1\ldots\alpha_L}(q^\perp_1)
\frac{\sqrt s\! +\!\hat P}{2\sqrt s}\;
\nn &&
R^{\alpha_1\ldots\alpha_L}_{\beta_1\ldots\beta_L}
V^{(i+)\mu}_{\beta_1\ldots\beta_L}(k^\perp)
u(k_1)  \varepsilon_\mu  \;,
\label{double11}
\ee
where the $k_1$ and $q_1$ are the momenta of the
nucleon in the initial and the final state,
$k^\perp = 1/2(k_1-k_2)^\perp$ and
$q_1^\perp = 1/2(q_1+q_2-q_3)^\perp$ are their
components orthogonal to the total momentum of
the first resonance $R_1$. Further, $s_{12}=(q_1+q_2)^2$ and
the factors $1/(2\sqrt s)$ and $1/(2\sqrt{s_{12}})$ are introduced
to suppress the divergency of the numerator of the fermion  propagators
at large energies.
The relative momentum $q_{12}^\perp$
is the component  of $q_1$ and
$q_2$ orthogonal to the total momentum $q_1+q_2$. It is given by:
 \be
q_{12\mu}^\perp=
\frac 12 (q_1-q_2)_\nu \left (g_{\mu\nu}-
\frac{(q_1+q_2)_\mu(q_1+q_2)_\nu}{(q_1+q_2)^2} \right) &&
\label{q12}
\ee
The vertex functions (\ref{sum_v1})-(\ref{sum_v2})
are given for the case when the nucleon wave function is placed on the
right-hand side of the amplitude. Therefore the  order of the
$\gamma$-matrices needs to be changed for the meson-nucleon vertices in
eq.(\ref{double1}).  

If the baryon $R_2$  has spin $1/2^-$ one
has to construct the vertex for decay of '+' states into a $0^-$
and a $1/2^-$ particle. However such operators coincide with the
operators for the decay of '-' states into a $0^-+1/2^+$ system.
Therefore
\be
&& A^{(i)}= \bar u(q_1)\tilde N^+(q^\perp_{12})\;
\frac{\hat q_1\! +\! \hat q_2\! +\! \sqrt {s_{12}}}{2\sqrt{s_{12}}}
\tilde N^-_{\alpha_1\ldots\alpha_L}(q^\perp_1)
\nn &&
F^{\alpha_1\ldots\alpha_L}_{\beta_1\ldots\beta_L}(P)
V^{(i+)\mu}_{\beta_1\ldots\beta_L}(k^\perp)
u(k_1)  \varepsilon_\mu =
\nn &&
\bar u(q_1) \;
\frac{\hat q_1\! +\! \hat q_2\! +\! \sqrt {s_{12}}}{2\sqrt{s_{12}}}
i\gamma_\nu \gamma_5 X^{(L+1)}_{\nu\alpha_1\ldots\alpha_L}(q^\perp_1)
\nn &&
F^{\alpha_1\ldots\alpha_L}_{\beta_1\ldots\beta_L}(P)
V^{(i+)\mu}_{\beta_1\ldots\beta_L}(k^\perp)
u(k_1)  \varepsilon_\mu  \;.
\label{double4}
\ee
In case of the photoproduction
with real photons, the $V^{(2+)\mu}_{\beta_1\ldots\beta_L}$ vertex is
reduced to $V^{(1+)\mu}_{\beta_1\ldots\beta_L}$, and can be omitted.

\subsubsection{The
$1/2^+,\;3/2^-,\;5/2^+\ldots$ states}

If a '-' state is produced in a $\gamma^* N$ interaction and then
decays into a pseudoscalar pion and $1/2^+$ baryon the
amplitude has the structure
\be
&& A^{(i)}\!=
\bar u(q_1)\tilde N^-(q^\perp_{12})\;
\frac{\hat q_1\! +\! \hat q_2\! +\! \sqrt {s_{12}}}{2\sqrt{s_{12}}}
\tilde N^-_{\alpha_1\ldots\alpha_{L-1}}(q^\perp_1)
\nn &&
F^{\alpha_1\ldots\alpha_{L-1}}_{\beta_1\ldots\beta_{L-1}}(P)
V^{(i-)\mu}_{\beta_1\ldots\beta_{L-1}}(k^\perp)
u(k_1)  \varepsilon_\mu =
\nn &&
\bar u(q_1)\;i\hat q^\perp_{12} \gamma_5 \;
\frac{\hat q_1\! +\! \hat q_2\! +\! \sqrt {s_{12}}}{2\sqrt{s_{12}}}
i\gamma_\nu\gamma_5
X^{(L)}_{\nu\alpha_1\ldots\alpha_{L-1}}(q^\perp_1)
\nn &&
F^{\alpha_1\ldots\alpha_{L-1}}_{\beta_1\ldots\beta_{L-1}}(P)
V^{(i-)\mu}_{\beta_1\ldots\beta_{L-1}}(k^\perp)
u(k_1)  \varepsilon_\mu  \;.
\label{double7}
\ee
If the intermediate baryon has spin $1/2^-$ then:
\be
&& A^{(i)}\!=
\bar u(q_1)\tilde N^+(q^\perp_{12})\;
\frac{\hat q_1\! +\! \hat q_2\! +\! \sqrt {s_{12}}}{2\sqrt{s_{12}}}
\tilde N^+_{\alpha_1\ldots\alpha_{L-1}}(q^\perp_1)
\nn &&
F^{\alpha_1\ldots\alpha_{L-1}}_{\beta_1\ldots\beta_{L-1}}(P)
V^{(i-)\mu}_{\beta_1\ldots\beta_{L-1}}(k^\perp)
u(k_1)  \varepsilon_\mu =
\nn &&
\bar u(q_1) \;
\frac{\hat q_1\! +\! \hat q_2\! +\! \sqrt {s_{12}}}{2\sqrt{s_{12}}}
X^{(L)}_{\alpha_1\ldots\alpha_{L}}(q^\perp_1)
\nn &&
F^{\alpha_1\ldots\alpha_{L-1}}_{\beta_1\ldots\beta_{L-1}}(P)
V^{(i-)\mu}_{\beta_1\ldots\beta_{L-1}}(k^\perp)
u(k_1)  \varepsilon_\mu \;.
\label{double41}
\ee
For photoproduction with real photons only amplitudes
with $V^{(1-)}$ and $V^{(3-)}$ vertex functions should be
taken into account.


\subsection{Photoproduction amplitudes for baryon
states decaying into a \boldmath $3/2$ \boldmath state and a
pseudoscalar meson}

Experimentally important is photoproduction of
resonances decaying into $\Delta(1232) \pi$  followed by a
$\Delta(1232)$ decay into a nucleon and a pion.

\subsubsection{The $1/2^-,\;3/2^+,\;5/2^-\ldots$ states
decaying into a meson with spin $0$  and a baryon with spin $3/2$}

The '+' states produced in a $\gamma^* N$
collision can decay into a pseudoscalar meson and an intermediate
baryon with spin $3/2^+$ in two partial waves. The amplitude
depends on indices $(ij)$ where index $(i)$ is related, as before, to
the partial wave in the $\gamma N$ channel while index $(j)$ is related
to the partial wave in the decay of the resonance into the spin $0$
meson and the $3/2$ resonance $R_2$.
\be
&& A^{(ij)}= \bar u(q_1)\;\tilde
N^+_\delta(q^\perp_{12}) F^\delta_\nu(q_1+q_2)\; \tilde
N^{(j+)\nu}_{\alpha_1\ldots\alpha_L}(q^\perp_1)
\nn &&
F^{\alpha_1\ldots\alpha_L}_{\beta_1\ldots\beta_L}(P)
V^{(i+)\mu}_{\beta_1\ldots\beta_L}(k^\perp)
u(k_1) \varepsilon_\mu  \;.
\label{double12}
\ee

If the intermediate baryon $R_2$
has $J^P=3/2^-$, the structure of the amplitude structure is
\be
&& A^{(ij)}=
\bar u(q_1)\;\tilde N^-_\delta(q^\perp_{12}) F^\delta_\nu(q_1+q_2)\;
\tilde N^{(j-)\nu}_{\alpha_1\ldots\alpha_L}(q^\perp_1)
\nn &&
F^{\alpha_1\ldots\alpha_L}_{\beta_1\ldots\beta_L}(P)
V^{(i+)\mu}_{\beta_1\ldots\beta_L}(k^\perp)
u(k_1) \varepsilon_\mu  \;.
\label{double15}
\ee


\subsubsection{The $1/2^+,\;3/2^-,\;5/2^+\ldots$ states
decaying into a $0^-$ meson and a $3/2^+$ baryon}

The amplitudes for '-' states
decaying into $0^-$ meson and  $3/2^+$ intermediate baryon are
\be
&& A^{(ij)}=
\bar u(q_1)\;\tilde N^+_\delta(q^\perp_{12}) \; F^\delta_\nu(q_1+q_2)\;
\tilde N^{(j-)\nu}_{\alpha_1\ldots\alpha_{L-1}}(q^\perp_1)
\nn &&
F^{\alpha_1\ldots\alpha_{L-1}}_{\beta_1\ldots\beta_{L-1}}(P)
V^{(i-)\mu}_{\beta_1\ldots\beta_{L-1}}(k^\perp)
u(k_1)  \varepsilon_\mu \;.
\label{double13}
\ee
and if the intermediate baryon $R_2$ has the quantum numbers $3/2^-$
\be
&& A^{(ij)}=
\bar u(q_1)\;\tilde N^-_\delta(q^\perp_{12}) \; F^\delta_\nu(q_1+q_2)\;
\tilde N^{(j+)\nu}_{\alpha_1\ldots\alpha_{L-1}}(q^\perp_1)
\nn &&
F^{\alpha_1\ldots\alpha_{L-1}}_{\beta_1\ldots\beta_{L-1}}(P)
V^{(i-)\mu}_{\beta_1\ldots\beta_{L-1}}(k^\perp)
u(k_1)  \varepsilon_\mu  \;.
\label{double14}
\ee

\section{t- and u-channel exchange amplitudes}

Meson exchange in the t-channel
plays an important role in both, in photoproduction and in
pion induced reactions. Especially at large energies this mechanism often
dominates. In the resonance region we expect that
production of baryon resonances in the s-channel dominates the
interaction, at least when neutral mesons are produced. Nevertheless the
t- and u-channel exchanges must be taken into account carefully.  %

The most straight forward parameterization of particle exchange
amplitudes is the exchange of Regge trajectories. For construction of a
cross-symmetrical amplitude it is convenient to use the variable
$$\nu=\frac 12 (s-u).$$

The amplitude for t-channel exchange can be written as
\be
A=g_1(t)g_2(t)\frac{1+\xi exp(-i\pi\alpha(t))}{\sin(\pi\alpha(t))}
\left (\frac{\nu}{\nu_0} \right )^{\alpha(t)} \;.
\ee
Here $g_i$ are vertex functions, $\alpha(t)$ is the function which describes
the trajectory, $\nu_0$ is a normalization factor (which can be taken to
be 1) and $\xi$ is the signature of the trajectory. The Pomeron, $f_0$
and $\pi$ exchanges have a positive signature while $\rho$, $\omega$
and $a_1$ exchanges have a negative one.

Accordingly, the Reggeon propagators can be written as
\be
R(+,\nu,t)=\frac{e^{-i\frac{\pi}{2}\alpha(t)}}
{\sin (\frac{\pi}{2}\alpha(t))}
\left (\frac{\nu}{\nu_0}\right )^{\alpha(t)}\;,
\nn
R(-,\nu,t)=\frac{ie^{-i\frac{\pi}{2}\alpha(t)}}
{\cos (\frac{\pi}{2}\alpha(t))}
\left (\frac{\nu}{\nu_0}\right )^{\alpha(t)}  \;.
\label{7}
\ee
where '+' and '-' indicate the signature of the Regge-trajectories.
To eliminate the poles at $t<0$ additional $\Gamma$-functions are introduced in
(\ref{7}). If the Pomeron trajectory is taken as $1.0+0.15t$
\cite{pom_slope}, negative $t$ poles are at $\alpha=0,-2,-4,\ldots$
and therefore
\be
\sin \left (\frac{\pi}{2}\alpha(t)\right )
\to
\sin \left (\frac{\pi}{2}\alpha(t)\right ) \;
\Gamma \left (\frac{ \alpha(t)}{2}\right ) \;.
\label{pomeron_gam}
\ee
For the pion trajectory $\alpha(t)=-0.014+0.72t$ \cite{pom_slope}, and
the negative poles are at $\alpha=-2,-4,\ldots$. Regularization
must be taken as
\be
\sin \left (\frac{\pi}{2}\alpha(t)\right ) \to
\sin \left (\frac{\pi}{2}\alpha(t)\right ) \; \Gamma \left (\frac{
\alpha(t)}{2} +1\right ) \;.
\label{12}
\ee
For $\rho$, $\omega$ and $a_1$
exchanges the negative poles start from $a=-1$ and therefore
\be
\cos \left (\frac{\pi}{2}\alpha(t)\right ) \to \cos \left
(\frac{\pi}{2}\alpha(t)\right )  \; \Gamma \left (\frac {\alpha(t)}{2}
+\frac 12\right )\, .
\ee

\subsection{Single meson photoproduction due to \boldmath $\rho$ \unboldmath and
\boldmath $\omega$ \unboldmath exchange}

In the following, the 4-vectors of the initial photon and proton are
denoted as $k_1$ and $k_2$  and 4-vectors of the final state nucleon
(e.g. proton) and the meson (e.g. pion) as $q_1$ and $q_2$ respectively
(see Fig.~\ref{tchn_s}).
The photon couples to the $\pi \rho(770)$ system in a
P-wave, and the corresponding amplitude for upper vertex is
\be
A_{upper}=\varepsilon_\mu
\rho_\alpha \epsilon_{\mu\alpha\beta\gamma}\; q_{2\beta}\;k_{2\gamma}\;,
\label{x1}
\ee
where $\rho_\alpha$ is the polarization vector of $\rho$-meson. \\
\begin{figure}[h]
\leftline{
\epsfig{file=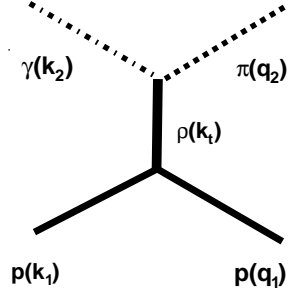,width=14.cm,clip=on}}
\caption{The t-channel exchange diagram
for single meson photoproduction}
\label{tchn_s}
\end{figure}
Another
vertex in this diagram describes the transition of the proton and the
$\rho$-meson into the final proton.
Such a transition has the same vertex structure as the
transition $\gamma^* N$ to a nucleon at the lower vertex:
\be
&&A_{lower}^{(i-)}=\bar u(q_1) V^{(i-)\mu}(k_1^{\perp}) u(k_1)
\rho_\mu\;,\qquad \qquad i=1,2
\nn
&&k_{1\mu}^{\perp}=
k_{1\nu}\Big ( g_{\mu\nu}-\frac{q_{1\mu} q_{1\nu}}{q_1^2}\Big )=
\frac 12(k_1-k_t)_{\nu}
\Big ( g_{\mu\nu}-\frac{q_{1\mu} q_{1\nu}}{q_1^2}\Big )\;.
\nn
\ee
Here $k_t = k_2 - q_2 = q_1-k_1$ is the $\rho$-meson momentum.
Summing over its polarizations yields
\be
\sum\limits_{polarization}\rho_\alpha\rho_\beta=
g_{\alpha\beta}-\frac{k_{t\alpha}k_{t\beta}}{k_t^2}  \;,
\ee
we obtain the following expression for the amplitude:
\be
A^{i-} =
\varepsilon_\mu \epsilon_{\mu\alpha\beta\gamma}\;
q_{2\beta}\;k_{2\gamma}
\bar u(q_1) V^{(i-)\alpha}(k_1^{\perp}) u(k_1) \;,\;\; i=1,2 \nn
\label{x3a}
\ee
The same amplitude structure corresponds also to $\omega$- exchange.

\subsection{Double meson photoproduction due to \boldmath $\rho$ \unboldmath and
\boldmath $\omega$ \unboldmath exchange}

Let us consider photoproduction of two meson (e.g. pions)
due to $\rho$ exchange  in t-channel
with a $1/2$ resonance in the intermediate state
(see Fig.~\ref{tchn_d}).
\begin{figure}[h]
\leftline{
\epsfig{file=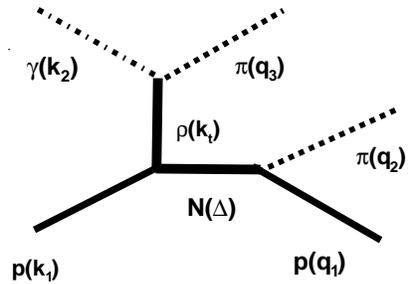,width=14.cm,clip=on}}
\caption{The t-channel exchange diagram for double meson
photoproduction reactions}
\label{tchn_d}
\end{figure}
In this
case we should add to eq.(\ref{x3a}) the $1/2$ propagator and
the vertex for decay of this resonance
into final meson and nucleon:
\be
&& A^{i\pm} = \varepsilon_\mu \epsilon_{\mu\alpha\beta\gamma}\;
q_{3\beta}\;k_{2\gamma}
\bar u(q_1)\tilde N^\pm(q^\perp_{12})\;
\frac{\hat q_1\! +\! \hat q_2\! +\! \sqrt {s_{12}}}{2\sqrt{s_{12}}}
\nn &&
V^{(i\pm)\alpha}(k_1^{\perp}) u(k_1)\;, \quad \quad i=1,2\;
\label{x4a}
\ee
{\rm with \qquad} $ k_{1\mu}^\perp=
k_{1\nu}\big ( g_{\mu\nu}-(q_1+q_2)_\mu (q_1+q_2)_\nu/s_{12}\big ) $. \\[+2ex] %
The definition of $q_{12}^\perp$ is given in (\ref{q12}).
The '-' amplitude corresponds to production of a $1/2^+$
intermediate state while the '+' amplitude corresponds to production of
a $1/2^-$ intermediate state.

Two-meson photoproduction due to
$\rho$ exchange in t-channel with a $3/2$ resonance in the intermediate
state can be easily obtained following the procedure given above. In
\be
&& A^{i\pm} = \varepsilon_\mu \epsilon_{\mu\alpha\beta\gamma}\;
q_{3\beta}\;k_{2\gamma}
\bar u(q_1)\tilde N^\pm_\xi(q^\perp_{12})\; \nn &&
F^\xi_\chi(q_1+q_2)
V^{(i\pm)\alpha}_\chi (k_1^{\perp}) u(k_1) \qquad i=1,2,3
\label{tex_delta}
\ee
the '-' amplitude corresponds to a $3/2^-$ intermediate state and
'+' amplitude to $3/2^+$ intermediate state.

The examples of other $t$-channel and u-channel exchange amplitudes
exchange amplitudes used in the analysis of the
single and double meson photoproduction  are given in Appendix D.

\section{The cross section for photoproduction processes}

The differential
cross section for production of two or more particles has the form:
\be
&&d\sigma=\frac{(2\pi)^4 |A|^2}{4\sqrt{(k_1k_2)^2-m_1^2m_2^2}}
d\Phi_n(k_1+k_2,q_1,\ldots,q_n)  \;,\nn
\ee
where $k_1$ and $k_2$ are momenta of the initial particles
(nucleon and $\gamma$ in the case of photoproduction) and $q_i$ are
momenta of final state particles. The $d\Phi_n(k_1+k_2,q_1,\ldots,q_n)$
is the element of the n-body phase volume given by
\be
d\Phi_n(k_1+k_2,q_1,\ldots,q_n)&=&\delta^4(k_1+k_2-
\sum\limits_{i=1}^n q_i)  \nn &&
\prod\limits_{i=1}^n \frac{d^3q_i}{(2\pi)^3 2q_{0i}}\;.
\ee
The photoproduction amplitude can be written as
\be
A=\varepsilon_\mu \bar u_i A_{\mu}
u_f \;,
\ee
where $\varepsilon_\mu$ is the $\gamma$ polarization vector
and $\bar u_i$ and $u_f$ are the bispinors of the initial
and final state nucleon. When the $\gamma$ and
nucleon polarization are not measured the amplitude squared is equal to
\be
|A|^2=
\frac 14 \sum\limits_{\alpha j k} A\;A^*=
\frac 14 \sum\limits_{\alpha j k}
\varepsilon^{\alpha}_\mu
\varepsilon^{\alpha}_\nu
\bar u^j_i A_{\mu} u^k_f \bar u^k_f A^{tr}_\nu u^j_i \;,\;
\ee
where one averages over the polarization of the initial and sums over
the polarization of the final state particles. $A^{tr}$ is the
hermitian conjugate amplitude.

For the unpolarized real photons:
\be
-\sum\limits_{\alpha}
\varepsilon^{\alpha}_\mu
\varepsilon^{\alpha}_\nu=g_{\mu\nu}^{\perp\perp}=
g_{\mu\nu}-\frac{P_\mu P_\nu}{P^2}-\frac{
k^\perp_\mu k^\perp_\nu}{k_\perp^2}
\ee
with $P=k_1+k_2$ and
\be
k^\perp_\mu=\frac 12 (k_1-k_2)_\nu g^\perp_{\mu\nu}=
\frac 12 (k_1-k_2)_\nu \left ( g^\perp_{\mu\nu}-\frac{P_\mu P_\nu}{P^2}
\right )    \nonumber
\ee
Let us remind that in the c.m.s. with the momentum of $\gamma$ being
parallel to the z-axis:
\be
g^{\perp\perp}_{\mu\nu}\qquad = \qquad
\left (
\begin{array}{cccc}
0 & 0 & 0 &0 \\
0 &-1 & 0 &0 \\
0 & 0 &-1 &0 \\
0 & 0 & 0 &0
\end{array}
\right )\;.
\ee

The bispinors of fermions with momentum $k_1$
summed over polarization are convoluted
(taking into account normalization (\ref{bisp_norm})) and yield
\be
\sum\limits_{j} u^j(k_1)\bar u^j(k_1)=m+\hat k_1
\ee
and therefore
\be
|A|^2=
\frac 14 g_{\mu\nu}^{\perp\perp}
Tr\left [(m+\hat k_1) A_{\mu} (m+\hat q_1)
A^{tr}_\nu \right ]\;.
\label{am2}
\ee
In case of a polarised target the density matrix of the fermion
propagator $(m+\hat k_1)$ must be changed to the polarization density
matrix:
\be
m+\hat k_1\to (m+\hat k_1)(1+\gamma_5\hat S_T) \;,
\ee
where the 4-vector $S_T$ is the polarization vector of the target
baryon ($S^2_T=-1$, $(k_1S_T)=0$).
If the polarisation of final baryon
is measured the density matrix of the
propagator $(m+\hat q_1)$ is substituted by:
\be
m+\hat q_1\to (m+\hat q_1)(1+\gamma_5\hat S_R)\;,
\ee
where the 4-vector $S_R$ is the polarization vector of the
final baryon ($S^2_R=-1$, $(q_1S_R)=0$).

When a $\gamma$ is linearly polarised along the x-axis
the  polarization vector is: $\varepsilon_\mu=(0,1,0,0)$ and
we do not need to average over two polarizations. Then
one has to change (\ref{am2}) by substituting
\be
\frac 12 g^{\perp\perp}_{\mu\nu}\qquad\to\qquad
\left (
\begin{array}{cccc}
0 & 0 & 0 &0 \\
0 & -1 & 0 &0 \\
0 & 0 & 0 &0 \\
0 & 0 & 0 &0
\end{array}
\right )
\ee
If one has a circular polarised beam
\be
\frac 12 g^{\perp\perp}_{\mu\nu}\qquad\to\qquad
\frac 12
\left (
\begin{array}{cccc}
0 & 0 & 0 &0 \\
0 & -1 & -i &0 \\
0 & i  & -1 &0 \\
0 & 0 & 0 &0
\end{array}
\right )
\ee

\section{Conclusion}

In the present paper the operator expansion
approach has been developed for the construction of amplitudes for pion-
and photon-induced reactions. The method is relativistically
invariant and can be easily applied to the construction of
amplitudes with multi-body final states.
For production of  pseudoscalar mesons the identity
of our amplitudes to the well known CGLN amplitudes is explicitly
shown. The formulas are given explicitly in the form used by the
Crystal Barrel at ELSA collaboration in the analysis of single and
double meson photoproduction.

\section*{Acknowledgments}
A.~V.~Anisovich and A.~V.~Sarantsev would like to thank the
Alexander von Humboldt foundation for generous support, A.A. for
a AvH fellowship and A.S. for the Friedrich-Wilhelm Bessel award.
U.~Thoma gratefully acknowledges a Emmy Noether grant by the
Deutsche Forschungsgemeinschaft.

\section*{Appendices}
\subsection*{A~~  Properties of Legendre polynomials}

The recurrent expression for Legendre polynomials is given by
\be
P_L(z)\ =\ \frac{2L-1}{L}\ z\ P_{L-1}(z)-\frac{L-1}{L}\ P_{L-2}(z)\;.
\ee
The first and second derivative of the Legendre polynomials can be
expressed as
\be
&&P'_{L}(z)\ =\ L\frac{P_{L-1}(z)-z\ P_L(z)}{1-z^2}\ =\ (L+1)
\frac{z\ P_L-P_{L+1}}{1-z^2}\  \;,\nonumber  \\
\ee
\be
P''_L(z)&=&\frac{2z\ P'_L(z)-L(L+1)\ P_L(z)}{1-z^2}\ \nn
&=&\
\frac{2P'_{L+1}(z)-(L+1)(L+2)\ P_L(z)}{1-z^2} \;.
\ee
Some other useful expressions given here for convenience are:
\be
&&P'_{L-1}=P'_L\ z-L\ P_L,\;\; P'_{L+1}=P'_L\ z+(L+1)\ P_L\;,
 \nn &&
P'_{L+1}-P'_{L-1}\ =\ (2L+1)P_L \;,
\nn &&
P''_{L+1}-P''_{L-1}\ =\ (2L+1)P'_L \;.\nonumber
\ee

\subsection*{B~~ Properties of angular momentum operators}

In the following we list useful properties of angular momentum
operators.

\begin{eqnarray}
&&X^{(n+1)}_{\mu\alpha_{1}\ldots\alpha_{n}}(q_{\perp})X^{(n)}_{\alpha_{1}
\ldots\alpha_{n}}(k_{\perp})
= \\ &&
\frac{\alpha_{n}}{n+1}
(\sqrt{k_{\perp}^2})^n (\sqrt{q_{\perp}^2})^{n+1} \biggl[
-\frac{k_{1\mu}}{\sqrt{k_{\perp}^{2}}}P'_n +
\frac{q_{1\mu}}{\sqrt{q_{\perp}^{2}}}P'_{n+1} \biggl] \;, \nonumber
\end{eqnarray}

\begin{eqnarray}
&& X^{(n)}_{\mu\alpha_{2}\ldots\alpha_{n}}(q_{\perp})X^{(n)}_{\nu\alpha_{2}\ldots\alpha_{n}}(k_{\perp})
= \nn &&
\frac{ \alpha_{n-1}}{n^2} (\sqrt{k_{\perp}^2})^n
(\sqrt{q_{\perp}^2})^{n}   \biggl[
g^{\perp}_{\mu\nu}P^{\prime}_{n-1}  - \biggl(
\frac{q_{\mu}^{\perp} q_{\nu}^{\perp}}{q_{\perp}^{2}} +
\frac{k_{\mu}^{\perp} k_{\nu}^{\perp}}{k_{\perp}^{2}} \biggl)
P^{\prime\prime}_{n}  +
\nonumber \\ &&
\frac{1}{2} \biggl( \frac{q_{\mu}^{\perp}k_{\nu}^{\perp}
 +
k_{\mu}^{\perp}q_{\nu}^{\perp}}{\sqrt{k_{\perp}^{2}}
\sqrt{q_{\perp}^{2}}} \biggl)
( P^{\prime}_{n} +2zP^{\prime\prime}_{n})
+ \\ &&
 \frac{2n-1}{2}  \biggl(
\frac{q_{\mu}^{\perp}k_{\nu}^{\perp} -
k_{\mu}^{\perp}q_{\nu}^{\perp}}{\sqrt{k_{\perp}^{2}}\sqrt{q_{\perp}^{2}}}
\biggl) P^{\prime}_{n} \biggl] \;, \nonumber
\label{x11}
\end{eqnarray}

\begin{eqnarray}
&& X^{(n+2)}_{\mu\nu\alpha_{1}\ldots\alpha_{n}}(q_{\perp})X^{(n)}_{\alpha_{1}\ldots\alpha_{n}}(k_{\perp})
= \nn &&
\frac{2}{3} \frac{ \alpha_{n}}{(n+1)(n+2)} (\sqrt{k_{\perp}^2})^n
(\sqrt{q_{\perp}^2})^{n+2}    \biggl(
\frac{X^{(2)}_{\mu\nu} (q_{\perp})}{q_{\perp}^{2}}
P^{\prime\prime}_{n+2} +
\nonumber \\  &&
\frac{X^{(2)}_{\mu\nu} (k_{\perp})}{k_{\perp}^{2}}P^{\prime\prime}_{n}-
\frac{3}{2} \frac{k_{\mu}^{\perp}q_{\nu}^{\perp}
+k_{\nu}^{\perp}q_{\mu}^{\perp} -\frac{2}{3}
g_{\mu\nu}^{\perp}(k^{\perp}q^{\perp})
}{\sqrt{k_{\perp}^{2}}\sqrt{q_{\perp}^{2}}} P^{\prime\prime}_{n+1} \biggl)\;, \nn
\end{eqnarray}

\begin{eqnarray}
&& X^{(n)}_{\alpha\gamma_{2}\ldots\gamma_{n}}(q_{\perp})
O_{\mu\beta_{2}\ldots\beta_{n}}^{\tau\gamma_{2}\ldots\gamma_{n}}
X^{(n)}_{\xi\beta_{2}\ldots\beta_{n}}(k_{\perp}) =
\nn &&
X^{(n)}_{\alpha\gamma_{2}\ldots\gamma_n}(q_{\perp})
\frac{g_{\tau\mu}}{n}
X^{(n)}_{\xi\gamma_{2}\ldots\gamma_n}(k_{\perp}) + \nonumber \\ &&
+
\frac{n-1}{n} X^n_{\alpha\mu\gamma_3\ldots\gamma_n}(q_{\perp})
X^{(n)}_{\xi\tau\gamma_{3}\ldots\gamma_n}(k_{\perp}) - \nn &&
\frac{2(n-1)}{n(2n-1)}
X^{(n)}_{\alpha\tau\gamma_{3}\ldots\gamma_n}(q_{\perp})
X^{(n)}_{\xi\mu\gamma_{3}\ldots\gamma_n}(k_{\perp}) \;.
\end{eqnarray}

\subsection*{C~~ Blatt-Weisskopf formfactors}

If a resonance with radius $r$ decays into two particle with (squared)
momentum $k^2$:
\begin{eqnarray}
k^{2}=\frac{(s-(m_1+m_2)^{2})(s-(m_1-m_2))^{2}}{4s} ,
\end{eqnarray}
where $s$ is  total energy and $m_1$ and $m_2$ are masses of the
final particles, then the first few expressions for
formfactors $F(L,k^2,r)$ are

\begin{eqnarray}
F(0,k^2,r) &=& 1 \;,\nn
F(1,k^2,r) &=& \frac{\sqrt{(x+1)}}{r} \;,\nn
F(2,k^2,r) &=& \frac{\sqrt{(x^2+3x+9)}}{r^2} \;,\\
F(3,k^2,r) &=& \frac{\sqrt{(x^3+6x^2+45x+225)}}{r^3} \;,\nn
F(4,k^2,r) &=& \frac{\sqrt{x^4+10x^3+135x^2+1575x+11025}}{r^4} \;, \nonumber
\end{eqnarray}
where $x=k^2 r^2$.
Remember that \\ $r(GeV^{-1})=r(fm)/(0.1973 (fm GeV))$.

\subsection*{D~~ Structure of amplitudes for t-channel and
u-channel exchanges}

\subsubsection*{D1~~ t-channel amplitudes}

For the photoproduction of a single neutral pion, $\rho$ and $\omega$
exchanges play a significant role. The exchange of a $\pi^0$ is forbidden
since the photon does not couple to a neutral pion.
When  charged pions are produced the
pion exchange diagram can play an important role. The
upper vertex function for pion exchange is
\be
A_{upper} &=&
\varepsilon_\mu \frac 12 (k_{2}+k_{t})_\nu
\Big (g_{\mu\nu}-\frac{q_{2\mu}q_{2\nu}}{m_\pi^2}\Big )= \nn &&
\varepsilon_\mu k_{t\nu} \Big (
g_{\mu\nu}-\frac{q_{2\mu}q_{2\nu}}{m_\pi^2}\Big ) \;.
\label{up_pi}
\ee
The lower vertex function is described by a $N\to \pi N$ transition.
Thus
\be
A=
\varepsilon_\mu k_{t\nu} \Big (
g_{\mu\nu}-\frac{q_{2\mu}q_{2\nu}}{m_\pi^2}\Big )
\bar u(q_1) N^{-}(k_1^{\perp}) u(k_1)  \;.
\label{app3_1}
\ee
Remember that for single meson production
\be
k_{1\mu}^{\perp}=
\frac 12(k_1-k_t)_{\nu}
\Big ( g_{\mu\nu}-\frac{q_{1\mu} q_{1\nu}}{q_1^2}\Big )=
k_{1\nu}\Big ( g_{\mu\nu}-\frac{q_{1\mu} q_{1\nu}}{q_1^2}\Big ) \;.\nn
\label{k1_single}
\ee

This expression can be easily extended to the case of double meson
photoproduction. If the intermediate baryon has spin $1/2$ one obtains
\be
A^\pm &=&
\varepsilon_\mu k_{t\nu} \Big (
g_{\mu\nu}-\frac{q_{2\mu}q_{2\nu}}{m_\pi^2}\Big ) \\&&
\bar u(q_1)\tilde N^\pm(q^\perp_{12})\;
\frac{\hat q_1\! +\! \hat q_2\! +\! \sqrt {s_{12}}}{2\sqrt{s_{12}}}
N^{\pm}(k_1^{\perp}) u(k_1)\;. \nonumber
\label{app3_2}
\ee
Here the '-' amplitude corresponds to a $1/2^+$ intermediate state,
the '+' to a $1/2^-$ state,
\be
k_{1\mu}^\perp=
k_{1\nu}\big ( g_{\mu\nu}-(q_1+q_2)_\mu (q_1+q_2)_\nu/s_{12}\big )\;,
\label{k1_double}
\ee
the definition of $q_{12}^\perp$ is given in eqn.(\ref{q12}) and
the notation of momenta is shown in Fig.~\ref{tchn_d}.

For an intermediate resonance with spin $J=L\pm 1/2$ the amplitude
structure reads
\be
A^{\pm} &=& \varepsilon_\mu k_{t\nu} \Big (
g_{\mu\nu}-\frac{q_{2\mu}q_{2\nu}}{m_\pi^2}\Big )
\bar u(q_1)\tilde N^\pm_{\alpha_1\alpha_2\ldots\alpha_n}(q^\perp_{12})\;
\nn &&
F^{\alpha_1\alpha_2\ldots\alpha_L}_{\beta_1\beta_2\ldots\beta_L}
(q_1+q_2)
N^{\pm}_{\beta_1\beta_2\ldots\beta_L} (k_1^{\perp}) u(k_1)
\label{app3_3}  \;.
\ee

The upper vertex for $\rho$-meson production due to pion exchange
has the following structure
\be
A_{upper} &=&
\varepsilon_\mu
\epsilon_{\mu\alpha\beta\gamma}\; \frac 12 (q_3-q_2)_\alpha
q_{2\beta}\;k_{2\gamma}=  \nn &&
\varepsilon_\mu
\epsilon_{\mu\alpha\beta\gamma}\; q_{3\alpha}q_{2\beta}\;k_{2\gamma}\;,
\ee
while the lower vertex has the same structure as the $\pi N$ scattering
amplitude. Therefore
\be
A=
\varepsilon_\mu
\epsilon_{\mu\alpha\beta\gamma}\; q_{3\alpha}q_{2\beta}\;k_{2\gamma}\;
\bar u(q_1) N^{-}(k_1^{\perp}) u(k_1)  \;.
\ee
Here $k_1^\perp$ is given by eq.(\ref{k1_single}).

The $\rho$ meson can also be produced by Pomeron or $f_0$
exchange. The upper vertex for such a case is
$\varepsilon_\mu \frac 12(q_3-q_2)_\mu $ and the amplitude is equal to
\be
A=\varepsilon_\mu \frac 12(q_3-q_2)_\mu
\bar u(q_1) N^{+}(k_1^{\perp}) u(k_1)  \;.
\ee

The next amplitude which we consider is the $f_0$ production due to
$\rho$ (or $\omega$) t-channel exchange. Such an amplitude has the
structure:
\be
A^{i-} =
\varepsilon_\mu \Big (g_{\mu\nu}-\frac{k_{t\mu}k_{t\nu}}{k^2_t}
\Big )
\bar u(q_1) V^{(i-)}_\nu(k_1^{\perp}) u(k_1) \;\; i=1,2 \nn
\ee

\subsubsection*{D2~~ u-channel amplitudes}

\begin{figure}[ht]
\epsfig{file=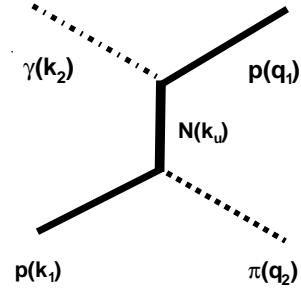,width=14.cm,clip=on}
\caption{The u-channel exchange diagram for photoproduction of single
mesons}
\label{uchn_s}
\end{figure}
Apart from  meson exchange
amplitudes (which we define as t-channel exchanges), mesons can be
produced from baryon exchange in the u-channel. An example of such a
diagram is given in Fig.~\ref{uchn_s}. For nucleon exchange, the vertex
for meson production (the lower vertex) is defined by
\be \bar u(k_u)
N^{-}(q_2^{\perp}) u(k_1) \;,\qquad \qquad k_u=k_1-q_2. \label{uchan1}
\ee
Here the $N^-$ vertex describes the production of a pseudoscalar
meson. Further,
\be
q_{2\mu}^\perp &=&
\left (g_{\mu\nu}-\frac{k_{1\mu} k_{1\nu}}{m_p^2}
\right )\frac 12 (q_2-k_u)_\nu = \nn &&
\left (g_{\mu\nu}-\frac{k_{1\mu} k_{1\nu}}{m_p^2}
\right ) q_{2\nu} .
\ee

\begin{figure}[h]
\epsfig{file=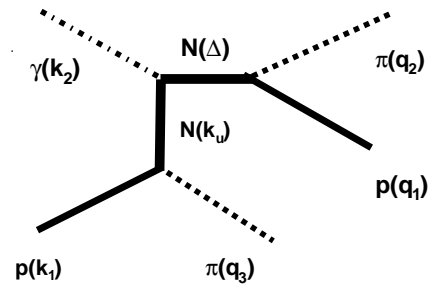,width=14.cm,clip=on}
\caption{A u-channel exchange diagram for production of a baryon
resonance in photoproduction of two mesons}
\label{uchn_d}
\end{figure}
If the reaction is induced by a meson the upper vertex has the same
structure
\be
\bar u(q_1) N^{-}(k_2^{\perp}) u(k_u)\;,
\label{uchan11}
\ee
where
\be
k_{2\mu}^\perp=
\left (g_{\mu\nu}-\frac{q_{1\mu} q_{1\nu}}{m_p^2}
\right )\frac 12 (k_2-k_u)_\nu=
\left (g_{\mu\nu}-\frac{q_{1\mu} q_{1\nu}}{m_p^2}
\right ) k_{2\nu} \nonumber
\ee
The angular dependent part of the amplitude for the nucleon exchange
diagram is
\be
A=\bar u(q_1) N^{-}(k_2^{\perp}) \frac{m_p+\hat k_u}{m_p^2-k_u^2}
N^{-}(q_2^{\perp}) u(k_1) \;.
\label{uchen_p1}
\ee
In the case of photoproduction the upper vertex is defined by
$V^{(i-)}_\mu$:
\be
A^i=\varepsilon_\mu
\bar u(q_1) V^{(i-)}(k_2^{\perp}) \frac{m_p+\hat k_u}{m_p^2-k_u^2}
N^{-}(q_2^{\perp}) u(k_1), \;\; i=1,2  \nn
\label{uchen_g1}
\ee

The production of $0^{++}$ states in double meson production
can be obtained from eqs.(\ref{uchen_p1},\ref{uchen_g1})
by replacing $N^-(q_2^{\perp})$ by $N^+(q_2^{\perp})$.

In the case when a baryon resonance with $J=L\pm 1/2$ is produced
in the intermediate state (see Fig.~\ref{uchn_d})
the amplitude for meson induced reaction has the structure
\be
&&A=\bar u(q_1)
N^{\pm}_{\alpha_1\alpha_2\ldots\alpha_L}(q_{12}^\perp)
F^{\alpha_1\alpha_2\ldots\alpha_L}_{\beta_1\beta_2\ldots\beta_L}
(q_1+q_2) \;,
\nn &&
 N^{\pm}_{\beta_1\beta_2\ldots\beta_L}(k_2^{\perp})
\frac{m_p+\hat k_u}{m_p^2-k_u^2}
N^{-}(q_3^{\perp}) u(k_1)
\ee
and for $\gamma^*$ induced reactions
\be
&& A^{i\pm}=\bar u(q_1)
V^{(i\pm)}_{\alpha_1\alpha_2\ldots\alpha_L}(q_{12}^\perp)
F^{\alpha_1\alpha_2\ldots\alpha_L}_{\beta_1\beta_2\ldots\beta_L}
(q_1+q_2)  \;,
\nn &&
 N^{\pm}_{\beta_1\beta_2\ldots\beta_L}(k_2^{\perp})
\frac{m_p+\hat k_u}{m_p^2-k_u^2}
N^{-}(q_3^{\perp}) u(k_1) \;.
\ee

\end{document}